\documentclass[a4paper,11pt]{article}

\usepackage{jinstpub} 
\usepackage{verbatim}
\usepackage[compact]{titlesec}

\usepackage{lineno}
\usepackage[version=4]{mhchem}
\usepackage{enumitem}

\makeatletter
\gdef\@fpheader{}
\makeatother

\title{\boldmath Observation of strong wavelength-shifting in the argon-tetrafluoromethane system}

\author[a]{P. Amedo,}
\author[a]{D. González-Díaz,}
\author[b]{F. M. Brunbauer,}
\author[a]{D. J. Fernández-Posada,}
\author[b]{E. Oliveri,}
\author[b]{L. Ropelewski}

\affiliation[a]{IGFAE, Universidade de Santiago de Compostela, Spain}
\affiliation[b]{CERN, Geneva, Switzerland}

\emailAdd{pablo.amedo.martinez@usc.es}

\abstract{We report the scintillation spectra of Ar/CF$_4$ mixtures in the range 210-800~nm, obtained under X-ray irradiation for various pressures (1-5~bar) and concentrations (0-100\%). Special care was taken to eliminate effects related to space charge and charge recombination, so that results can be extrapolated following conventional wisdom to those expected for minimum ionizing particles under the typical electric fields employed in gaseous instrumentation. Our study sheds light into the microscopic pathways leading to scintillation in this family of mixtures and reinvigorates the prospects of use in next-generation scintillation-based chambers.}


\begin{document}
\maketitle
\flushbottom

\section{Introduction}

Since their introduction in 1974 \cite{Nygren}, time projection chambers (TPCs) have proved to be one of the most effective ways of detecting particles and reconstructing their trajectories. The versatility of these devices, being compatible with $B$-fields, allowing readout flexibility and a wide range of density media (from some 10's of mbar up to 10's of bar, liquid or even solid phase), makes them the perfect tool to study many different phenomena in particle physics \cite{DiegoReview}. 

Gas-based TPCs commonly operate with admixtures of noble gases with some molecular species, chiefly CH$_4$, i-C$_4$H$_{10}$, CF$_4$ or CO$_2$. TPCs make use of these additives to reduce the spatial spread and collection time of the primary ionization, minimize photon and ion feedback and, in general, to attain a greater stability. Among them, CF$_4$ exhibits some particularly interesting properties such as intense and broadband scintillation in the range 150-750~nm under primary \cite{Pansky, CF4_alphas_Moro, CF4_alphas_Moro_field} and secondary (field-assisted) \cite{CF4_seco_Moro, Fraga_GEM, Diego_Vienna} particle excitation, and very low electron diffusion \cite{Christophorou_diff}.
The VUV-visible scintillation yields induced by $\alpha$ particles in pure CF$_4$ are found in the range $1000$-$3000$~ph/MeV \cite{Pansky, CF4_alphas_Moro, Rumore, Lehault}, optical gains well above $10^4$ have been reported in CF$_4$-based mixtures in \cite{Diego_Vienna}, while diffusion coefficients have been shown to remain at the thermal limit up to pressure-reduced drift fields as high as 1~kV/cm/bar \cite{Christophorou_diff}.

Based on the aforementioned observations, CF$_4$ by itself makes an interesting TPC gas (and has been used to that aim before, e.g. \cite{JOCELYN, Compton}). In fact, CF$_4$ either pure or admixed with other elements is of great contemporary interest to the optical imaging of rare processes in low-pressure gases \cite{Migdal, Betta1}. Ar/CF$_4$ admixtures, in particular, have been pioneered by the Fraga\&Fraga group at Coimbra already in the 00's for optical imaging \cite{Fraga1, Fraga2}, and revived recently in an optical-TPC demonstrator equipped with a triple-stack of gas electron multipliers (GEMs), \cite{Diego_Vienna, FlorianOTPC}. These works consistently showed a higher optical gain compared to pure CF$_4$, with indirect evidence for wavelength-shifting reactions between Ar states and the CF$_4$ scintillation precursors. Besides the enhanced performance of Ar/CF$_4$ mixtures for GEM operation, argon is considerably more cost-effective and environment-friendly than CF$_4$. Compared to a traditional wavelength-shifter like N$_2$, main advantages of CF$_4$ are its strong scintillation in the visible range together with a much lower electron diffusion, potentially allowing sharper and brighter tracks from CMOS and CCD cameras, e.g, when instrumenting optical TPCs in the field of nuclear physics \cite{2protonPomorski, Zimmerman_thesis}.



Argon has another characteristic relevant to modern instrumentation: it is the element of choice of the DUNE experiment, where it acts simultaneously as target and detection medium both at its far and near detector complexes \cite{DUNE_TDR}. Specifically, an argon-rich high pressure TPC capable of reconstructing low-energy hadrons (down to 10's of MeV, at least) has been proposed by the collaboration \cite{NDGAr}. It is called to be the first detector to ever record neutrino interactions in a sparse medium, with $4\pi$ coverage and broad particle identification (PID) capabilities. In this context, enabling time-tagging through the primary scintillation produced in neutrino interactions, while preserving the argon medium as pure as possible (to avoid parasitic neutrino interactions), is the subject of ongoing investigations \cite{We1, We2}. Time tagging is an essential asset in the study of neutrino oscillations with TPCs as it is used for spill-assignment, absolute estimate of the drift distance and time-of-flight determination of the emerging particles \cite{ND_CDR}.

With this in mind, we performed a systematic study of the primary scintillation in the Ar/CF$_4$ system down to trace-amounts of the molecular additive, in order to better understand its wavelength-shifting capabilities. For that, a spectroscopic analysis was carried out under X-ray irradiation at varying pressures and CF$_4$ concentrations, at electric fields and ionization densities for which space charge and charge/light recombination effects are negligible. Following conventional wisdom (e.g., \cite{REF-ICRU, Micro_Diego}), measurements in these conditions should represent a good approximation to the scintillation by minimum ionizing particles, a typical metric for characterizing the response of a particle detector. The present work is structured as follows: in section \ref{Setup} the experimental setup and procedures are described, section \ref{results} compiles the scintillation spectra of the pure gases and Ar/CF$_4$ admixtures; in section \ref{Kinetic} we present a minimalistic kinetic model that describes the observations to good accuracy, and we finally end with a comparison with previous results and a summary of our main conclusions in \ref{conc}.



\section{Experimental setup} \label{Setup}

\begin{figure}[h!!!]
\centering 
\includegraphics[width=1\textwidth,origin=c,angle=0]{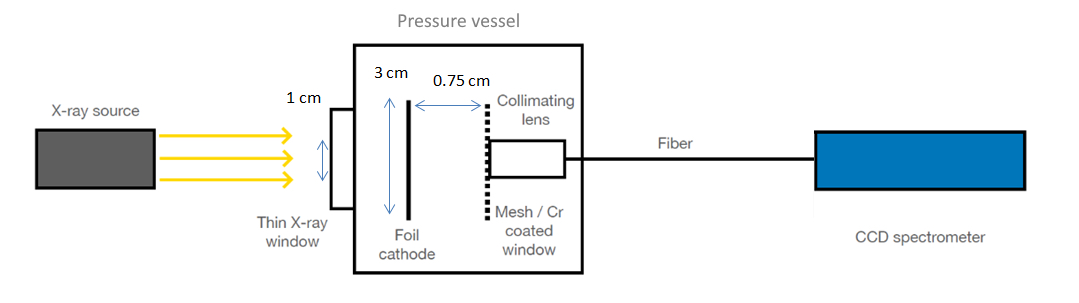}
\caption{Schematic drawing of the experimental setup used to conduct the measurements. The x-ray tube is shown to the left, the high-pressure vessel housing the electrifiable scintillating cylinder is shown at the middle. The spectrophotometer is shown to the right.}
\label{Chamber}
\end{figure}

Figure \ref{Chamber} shows a schematic drawing of the experimental setup. Measurements were performed on a CF63 aluminum-cube serving as a vessel, irradiated with X-rays from a copper tube at 40~kV. The chamber had an entrance window of 1~cm-diameter made of a thin aluminum foil of 50~{\textmu}m thickness which was facing the tube. Inside the chamber, an electrifiable cylindrical volume was placed, with 3~cm in diameter and 0.75~cm in height. Its upstream electrode served as a cathode and was made from the same foil as the window. A semitransparent Cr-mesh served as the anode, evaporated on top of a collimating lens (OceanOptics 74-UV) leading to a multi-mode optical fiber (UV-VIS, 600~µm core) and finally coupled to an OceanOptics FX UV-VIS CCD spectrometer sensitive in the 210-800~nm range. The photon spectrometer was calibrated using a lamp with reference light sources for the UV and the visible regions, coupled to the anode mesh. Both calibrations were merged at around the 300~nm mark.

Upon excitation and ionisation of the gas, electrons and ions were collected by means of an uniform electric field, the current being read at the anode with a Keithley picoamperemeter (model 6487). The maximum of the X-ray bremsstrahlung spectrum, when accounting for the absorption in the materials interposed up to the ionization region, was estimated to be at around 12~keV, a characteristic energy for which the X-ray mean free path is 14~cm in argon and 62~cm in CF$_4$, in standard conditions \cite{NIST_XCOM}. Even for argon at the highest pressures employed in our measurements (5~bar) the mean free path is as large as 2.8~cm, leading to $\pm10\%$-level variations within the ionization volume. The size of the ionization cloud ($\sigma$) caused by the tortuous trajectory of the ejected photoelectron amounts to a mere 0.25~mm$/P [\textnormal{bar}]$ in argon (see e.g. \cite{Azevedo}), much smaller than the chamber dimensions. The additional spread stemming from electron diffusion along a 0.75~cm drift-path, when considering electric fields at full charge-collection, can increase the above figure up to 1.6~mm in the radial direction (Pyboltz, \cite{Pyboltz}). This situation corresponds to pure Ar at 1~bar, with other conditions involving yet smaller charge spreads by roughly a factor of $1/\sqrt{P[\textnormal{bar}]}$ as the pressure increases, and up to another factor of ten as CF$_4$ concentration increases. Overall, inside the collimated region, ionization can be thus regarded as uniform throughout these measurements, for practical purposes.

\begin{figure}[h!]
\centering 
\makebox[\linewidth]{\includegraphics[scale=0.16]{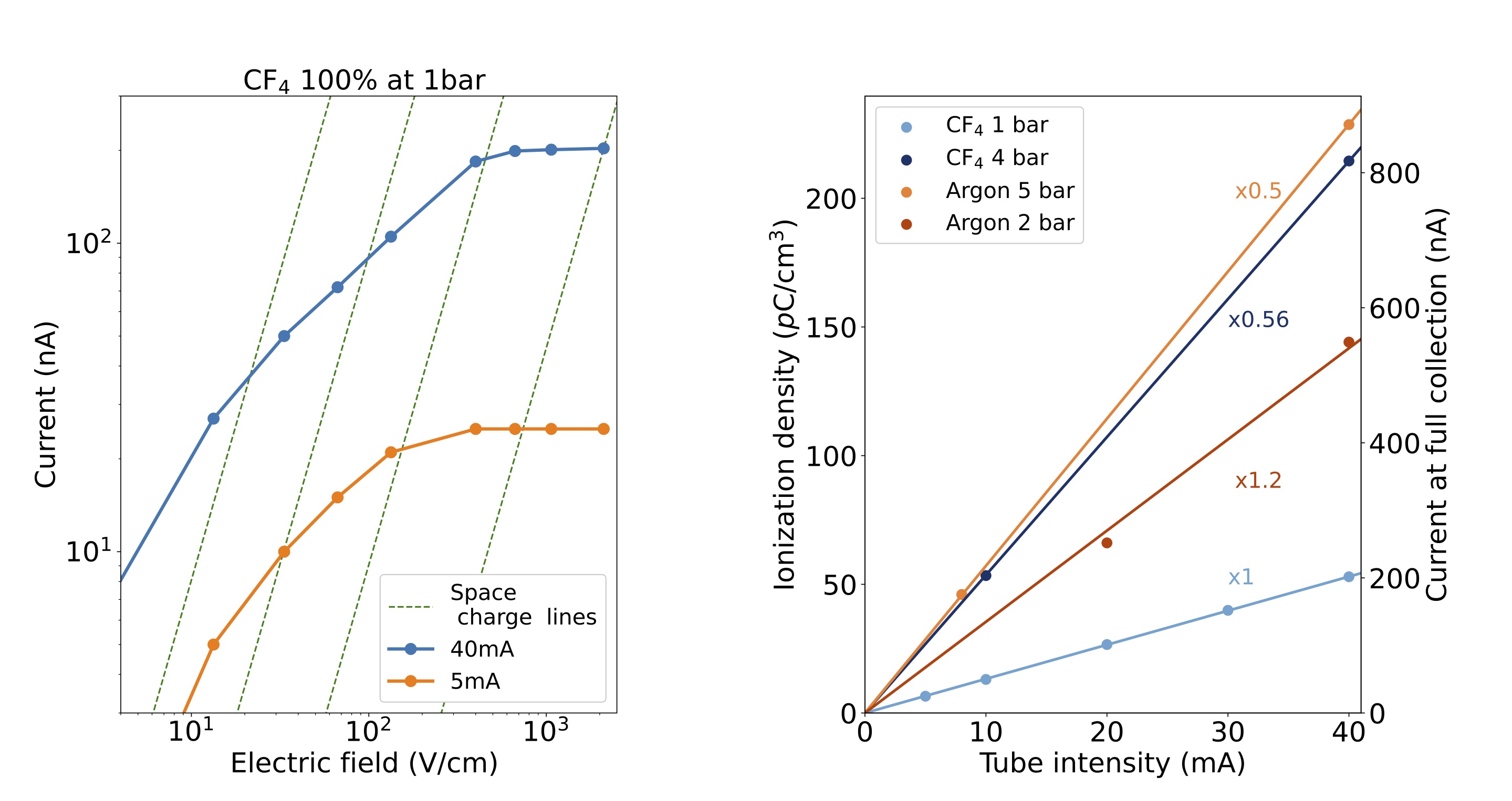}}
\caption{Exemplary scans in electric field and X-ray intensity. Left: current at the anode of the scintillation cell as a function of the applied electric field up to reaching full-collection, for different currents of the X-ray tube (pure CF$_4$ at 1~bar). Green lines represent identical space charge conditions (i.e., same degree of field distortion by positive ions) according to the parameter $\alpha$ introduced in  \cite{Space_charge_Mcdonald} and discussed in section \ref{EF_section} of the appendix. Right: current at full charge-collection (right axis) and associated ionization density (left axis) as a function of the current of the X-ray tube for CF$_4$ at 1 and 4 bar (light and dark blue) and pure argon at 2 and 5 bar (red and orange). Each of them was fitted to a proportional trend. (Given the non-linear map between the current and the ionization density, each data series includes an additional multiplicative factor in order to transform the right-axis value into the correct one)}
\label{CF4_1bar_currents}
\end{figure}

To exclude space charge and charge-recombination effects, data was taken at no field and at a field high enough to guarantee full charge-collection (Fig. \ref{CF4_1bar_currents}-left). Along this line, additional measurements were performed for different X-ray intensities too (e.g., Fig. \ref{CF4_1bar_currents}-right).
Under the assumption of uniform irradiation within the collimated region of the scintillation cell (of area $A=\pi* 0.5^2$ cm$^2$), the positive-ion space-charge density ($q_e\! \cdot \!dN/dV|_{ion}$) relates to the steady-state current at full collection through:

\begin{equation}
I_{sst} = 2 \left(q_e \frac{dN}{dV}\right)_{ion} \!\!\!\!\!\!\cdot A \cdot \mu \cdot E
\label{ion_density}
\end{equation}
with $q_e$ being the electron charge, $\mu$ the ion mobility,\footnote{A discussion on the inputs to the ion mobilities is given in the appendix.} $E$ the electric field, and the factor 2 accounts for the equal sharing of current between ions and electrons. Accordingly, the positive-ion space charge ranged in these measurements from 6~pC/cm$^3$ (for pure CF$_4$ at around 1~bar and the lowest X-ray intensity) up to 225~pC/cm$^3$ (for either pure Ar or CF$_4$ at around 5~bar, and the highest X-ray intensity). At 1~bar these values are about a factor of 4 below those employed in earlier measurements performed under $\alpha$ particles in \cite{CF4_alphas_Moro_field}, and reported to be recombination-free. Even the highest pressures explored in this work barely exceed the ionization densities studied earlier, which leads us to believe that recombination is negligible in present conditions. The good proportionality observed in Fig. \ref{CF4_1bar_currents}-right for different intensities of the X-ray tube adds further support to this.\footnote{A detailed argumentation on the absence of recombination in these measurements can be found in the appendix.}

In order to exclude any space charge effect from the positive ions, the analysis procedure sketched in \cite{Space_charge_Mcdonald} was applied. It follows, as  discussed in appendix (section \ref{EF_section}), that field distortions once the current reached saturation (full collection) were typically at the 5$\%$-level or below (with a maximum field distortion of 15$\%$) during the measurements. The resulting iso-space-charge lines (green dashed, in Fig. \ref{CF4_1bar_currents}-left) suggest that space charge is the main variable driving the current vs field behaviour. In the absence of space charge, the extent of fringe fields inside the chamber were evaluated through an electrostatic simulation. Results obtained with the COMSOL Multiphysics\textsuperscript{\textregistered} package \cite{COMSOL} indicate that the field is uniform within the ionization region, for practical purposes.\footnote{For details on the electric field calculations, including space charge, the reader is referred to the appendix.} 

Prior to the measurements, the chamber was pumped down to $10^{-4}$~mbar. Ar/CF$_4$ mixtures were studied at a volume fraction of 100/0, 99.9/0.1, 99.8/0.2, 99.5/0.5, 99/1, 98/2, 95/5, 90/10 and 0/100, and pressures from 1 to 5~bar. The purity of the bottles was 4.5 (CF$_4$) and 6 (Ar), so the overall purity of the studied mixtures was between 5.5 and 6 (i.e., 1-3~ppm contamination). The chamber was first filled with CF$_4$ until the desired partial pressure, using two pressure/vacuum gauges, namely a Pfeiffer Vacuum PCR 280 Pirani/Capacitance and an MKS pressure transducer, for the readings. Afterwards, argon was admixed. Gas circulation was dimmed unnecessary for proper mixing, as the concentration could be verified by sampling the gas into a residual gas analyzer and waiting for the ratio of the pressures of the species to stabilize. This agrees with the notion that the forced flow of argon gas, being dominant by at least a factor 10 in volume, drives the mixing in such a small chamber. For the lowest CF$_4$ concentrations, that would be limited by the accuracy of the sensor, the filling was done at high pressure and diluted until the target concentration was achieved. Deviations from the target CF$_4$ concentrations [0, 0.1, 0.2, 0.5, 1, 2, 5, 10, 100]$\%$ were quantified through a linear fit and associated uncertainties. Although the target values will be used as plot descriptors in the following, the calibrated values will be used when presenting systematics as well as for model fitting.\footnote{For details on the calibration procedure, the reader is referred to the appendix.}

Purity was monitored continuously with a residual gas analyzer (RGA) coupled to the main system through a leak valve. The RGA region was kept at a constant pressure of 10$^{-5}$ mbar throughout the measurements by adjusting the leak-valve opening. The main impurities in the system were H$_2$O, O$_2$ and N$_2$(CO) and their concentrations were estimated to be below 1000 H$_2$O ppm, 15 O$_2$ ppm and 200 N$_2$ ppm, being the sensitivity limited by the RGA background. These upper limits, as well as the scintillation yields, showed little variation with time, for a time span of hours. Even if we were to take them as representing the actual concentrations, it has been shown in \cite{Margato} that N$_2$ concentrations as high as 4\% are needed to quench CF$_4$ scintillation by a factor 2 (at 1~bar). Although 1000~ppms (0.1\% per volume) might arguably compete with Ar-CF$_4$ transfers at about the same CF$_4$ concentration (the lowest one used in our measurements), the phenomenological model introduced later in text does not show any strong deviation for that case. These observations, together with the nominal purity of the bottles, the use of low-outgassing materials for chamber assembly and the stability of the scintillation yields with time, suggest that the impact of impurities is of little relevance to the results presented in this work.

The final scintillation spectrum was divided by the current at full collection, and by the average energy to create an electron-ion pair ($W_I$). The latter was taken from the directly-measured values in \cite{Reinking}, except in the range [0-1]\% CF$_4$ where a simple linear interpolation was used. An absolute normalization was not attempted and thus the spectrum is hereafter expressed in yield/eV [a.u.]. As no significant contribution from recombination or space charge was found in present data, the standard deviation of measurements performed for different X-ray intensities has been used to estimate the uncertainty. This accounts for any residual recombination effect as well as systematic errors that may be present in the measurements.



\begin{figure}[h!!]
\makebox[\linewidth]{\includegraphics[scale=0.16]{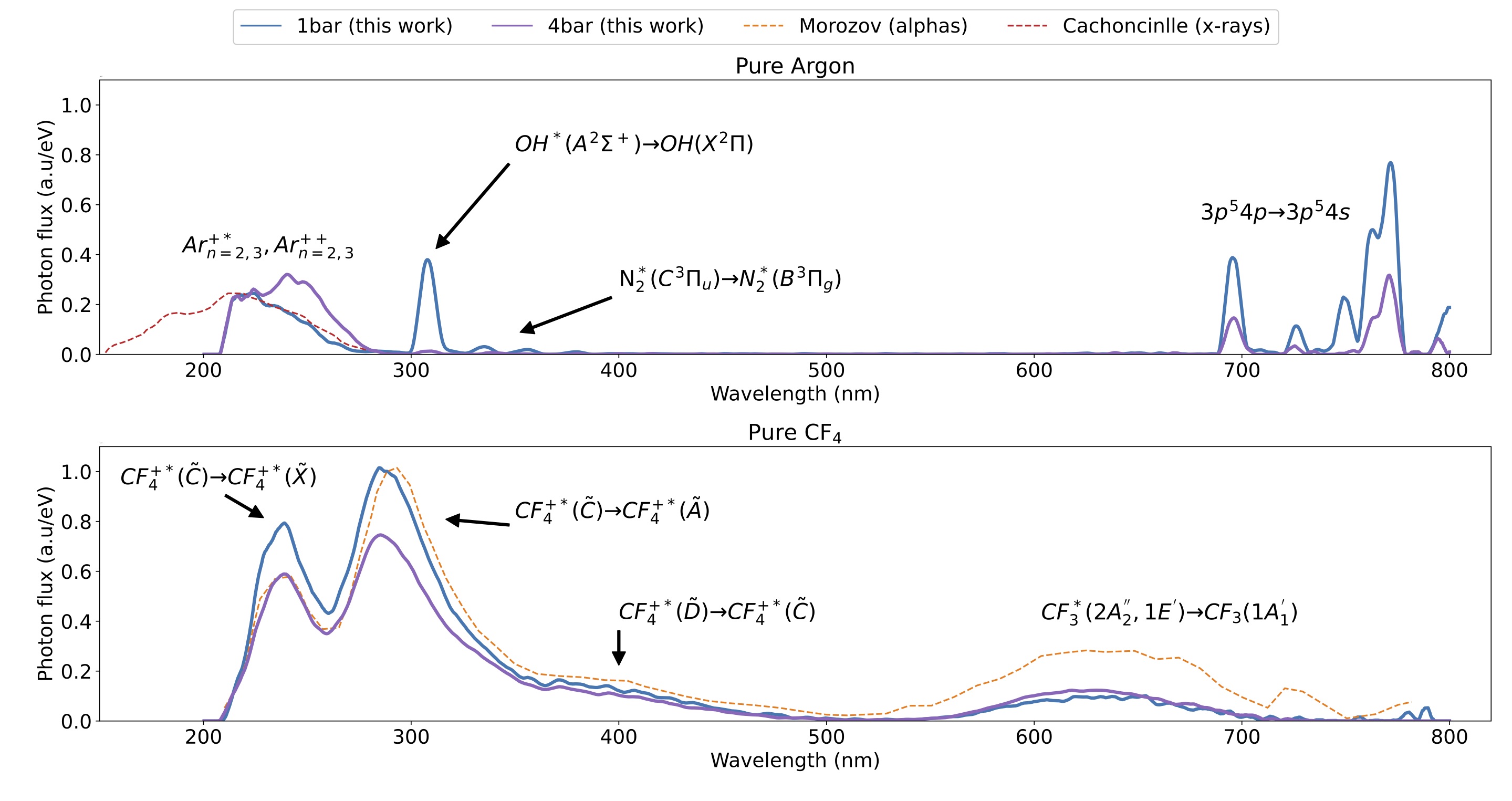}}
\caption{Emission spectra per eV of released energy for pure argon (upper plot) and pure CF$_4$ (lower plot) at 1 and 4 bar (blue, purple), obtained at zero field. No significant dependence with the intensity of the X-ray tube was observed, suggesting that measurements are recombination-free. For pure CF$_4$, where full charge-collection could be reached before the onset of secondary scintillation, no dependence with the electric field was observed either. The spectra of argon's $3^{\textnormal{rd}}$ continuum obtained in \cite{Cachoncille} also with X-rays (red-dashed) and the one of CF$_4$ obtained in \cite{CF4_alphas_Moro} with $\alpha$-particles (orange-dashed) are shown for comparison. A global normalization was imposed by setting to 1 the maximum of the 290 nm peak in pure CF$_4$ at 1 bar.}
\label{Pure_gases_spectra}
\end{figure}

\section{Results} \label{results}

\subsection{Pure gases}

Figure \ref{Pure_gases_spectra} shows the scintillation spectra of Ar (top) and CF$_4$ (bottom) at pressures of 1~bar (blue) and 4~bar (purple). Bands that are easily identifiable are the ones of the 3$^{\textnormal{rd}}$ continuum of argon (160-280~nm) \cite{Cachoncille} and the (210-500~nm) and (550-750~nm) ones of CF$_4$ \cite{CF4_alphas_Moro}. The visible band centered at around 630 nm has been attributed earlier to the transition CF$_3^*(2A_2^{''}, 1E^{'})\rightarrow$ CF$_3 (1A_1^{'})$ \cite{Suto_IV,Lee,Suto_III} (not being assigned unequivocally to either the $2A_2^{''}$ or the $ 1E^{'}$ states). The overlapping UV bands can be attributed to the CF$_4^{+,*}$ ion, emitting from its $\tilde{C}, \tilde{D}$ states. Transitions $\tilde{C} \rightarrow \tilde{X}, \tilde{A}$ can be naturally assigned to the peaks centered around 230 nm and 290 nm \cite{Lambert,Harshbarger,Zhang} while the transition at 364~nm may be assigned to $\tilde{D} \rightarrow \tilde{C}$. Another prominent UV band at around 260~nm has been observed before, e.g., under excitation within low-energy electron avalanches \cite{Fraga_GEM}, and can be assigned to the transition CF$_3^*(2A_1^{'})\rightarrow$ CF$_3 (1A_2^{''})$ \cite{Suto_IV,Lee}; it is however hidden in present conditions under the CF$_4^{+,*}$ emission. The small decrease of the yields in the UV bands as a function of pressure has been observed before in \cite{CF4_alphas_Moro} and might be naturally attributed to self-quenching. A comparison with data from \cite{CF4_alphas_Moro} obtained at 1~bar under $\alpha$-particle irradiation is shown in Fig. \ref{Pure_gases_spectra} (orange, dashed), arbitrarily normalized to the 290~nm peak. 

\begin{figure}[h!]
\makebox[\linewidth]{\includegraphics[scale=0.17]{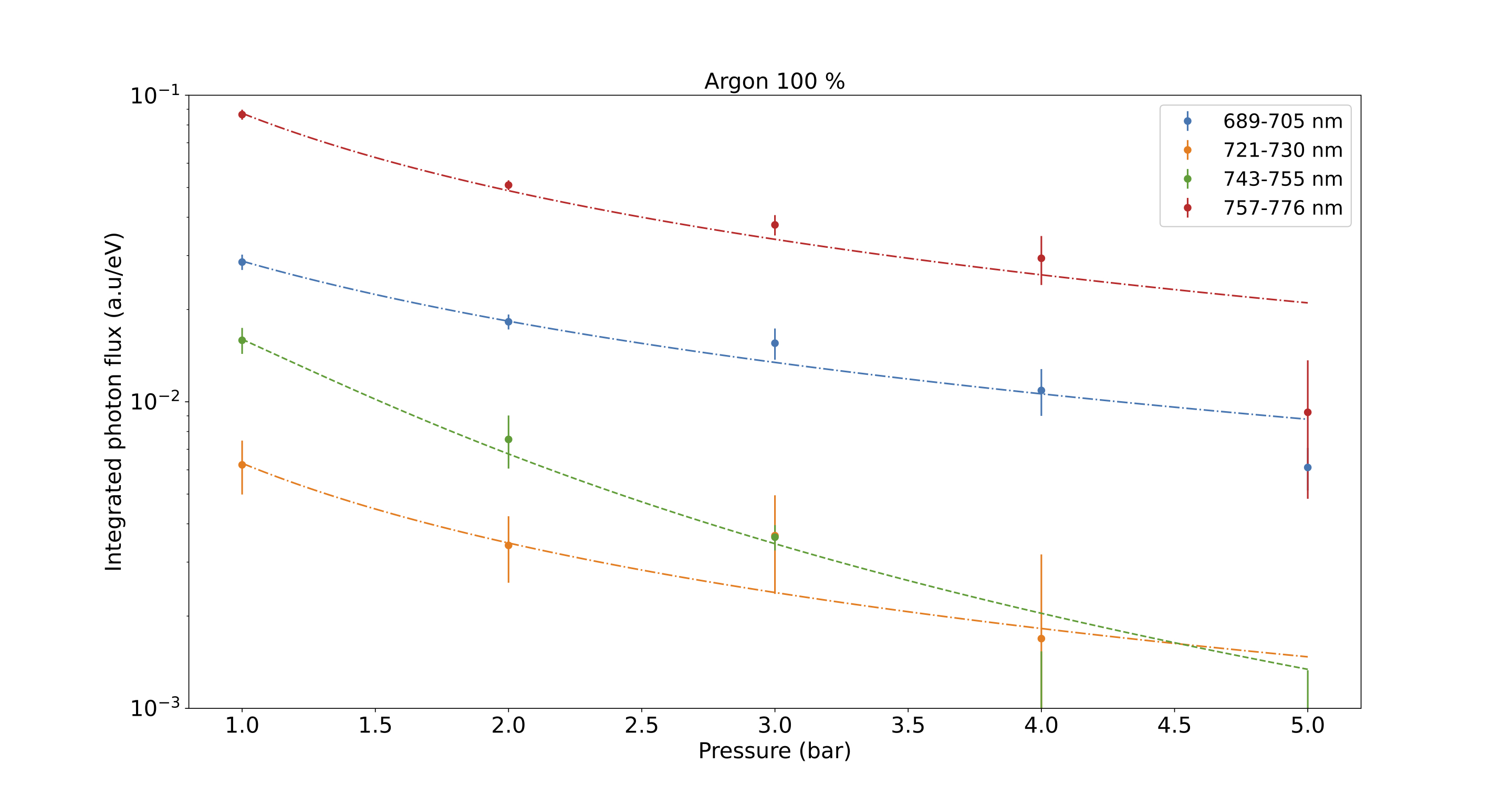}}
\caption{Integrated yields for different argon near-infrared peaks as a function of pressure. Dashed-dotted lines represent a fit to a 2-body self-quenching law while the dashed-green line follows from three-body self-quenching.}
\label{Peak_Yields_0V_40mA}
\end{figure}

Concerning Ar, the $3^{\textnormal{rd}}$ continuum (cut by the spectrometer bandwidth below 210~nm) agrees in shape with earlier X-ray measurements from \cite{Cachoncille}, increasing the yield on its blue-wing as the pressure increases, qualitatively in agreement with that work too (red dashed-line). In the near-infrared region the main lines located at 696, 727, 750, 763 and 772 nm can be clearly identified, corresponding to transitions between the $3p^54p$ and $3p^54s$ multiplets \cite{Wiese}. 
Their associated yields seem to be dominated by 2-body collisional self-quenching, thus approximately following a $\sim 1/(a+bP)$ trend, except in the case of the 750 nm peak where a $\sim 1/(a+bP^2)$ trend is observed instead (Fig. \ref{Peak_Yields_0V_40mA}). The strong suppression observed as a function of CF$_4$ concentration suggests that argon IR-yields will be subdominant in high pressure applications and/or as soon as a molecular additive is added.

The presence of impurities can be derived from the peak at around 310~nm, corresponding to the OH$^*(A^2\Sigma^+)\rightarrow$ OH$(X^2\Pi)$ transition \cite{Muller,OH_reco}. It can be attributed to charge transfer between Ar$^+$ and H$_2$O$^+$, following dissociative recombination to populate OH$^*$ \cite{OH_reco}. Even if barely visible, N$_2$ peaks at around 335, 355 and 380~nm are present in argon too, as expected from the transfer reactions identified in \cite{Takahashi}.

\subsection{Ar/CF$_4$ mixtures}

\begin{figure}[h]
\makebox[\linewidth]{\includegraphics[scale=0.18]{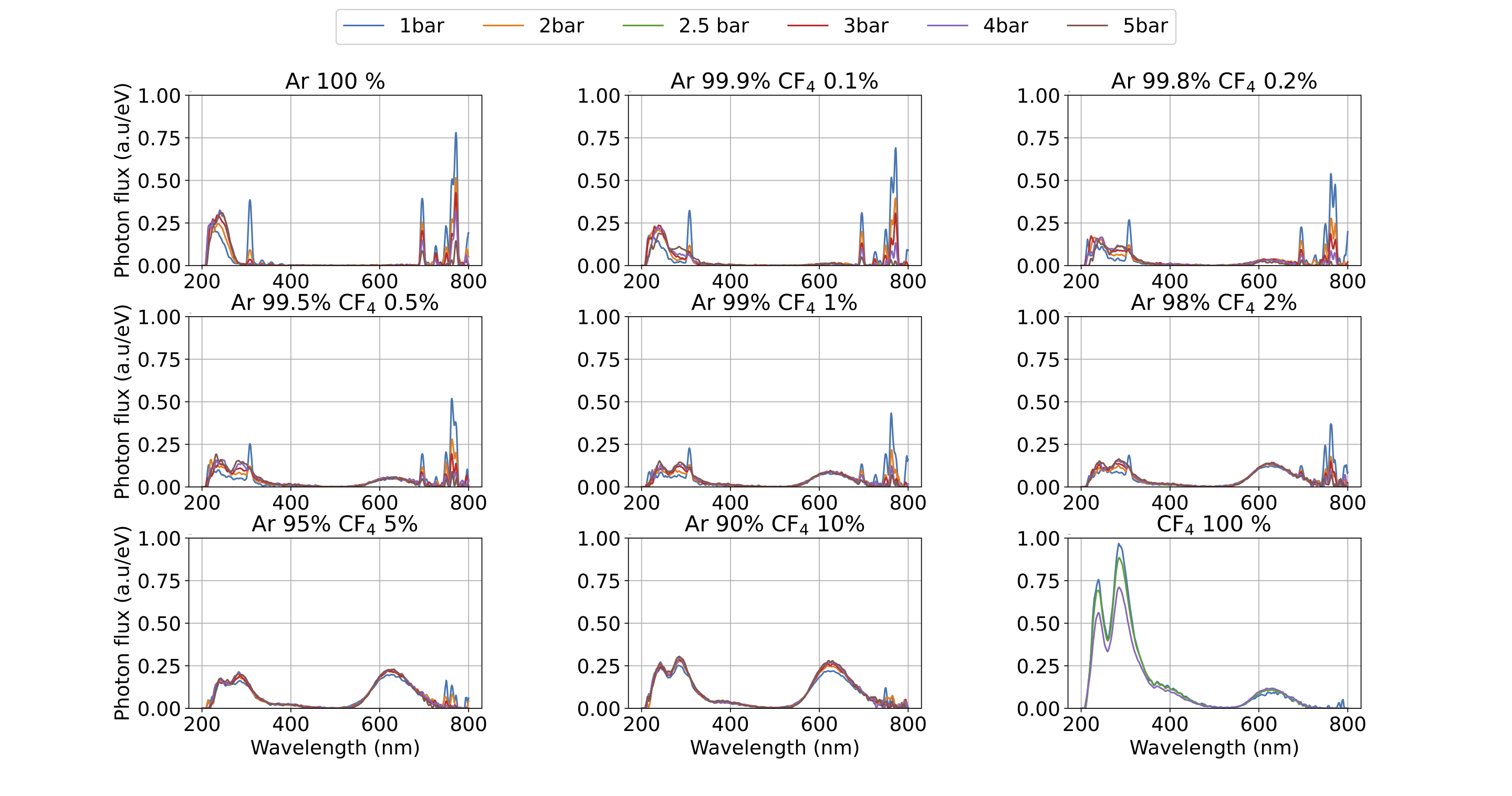}}
\caption{Emission spectra per eV of released energy for different concentrations of Ar/CF$_4$, at zero field. Measurements are expected to be free from recombination effects, as discussed in text. A global normalization was imposed by setting to 1 the maximum of the 290 nm peak in pure CF$_4$ at 1 bar. The exact CF$_4$ concentrations (after calibration) can be found in table \ref{CF4_RGA}. }
\label{Spectra_0V_40mA}
\end{figure}

Figure \ref{Spectra_0V_40mA} compiles the spectra for different Ar/CF$_4$ admixtures and pressures. Although they were obtained at zero field, no significant dependence with the X-ray intensity or electric field was observed, demonstrating the absence of recombination effects. This was generally the case except below 1\% CF$_4$, conditions for which the energy of the ionization electrons is high enough to cause neutral bremsstrahlung radiation (NBrS) during their drift \cite{Buzu, PRX, NBRS_JINST} at the fields required for full charge-collection, thus complicating the interpretation. Exemplary, for the full-collection field of ~2900 V/cm in pure Ar at 1~bar, NBrS would amount to about $\sim0.015$~ph/eV in the region 210-800~nm \cite{NBRS_JINST}. Given its flat nature, NBrS easily overwhelms the $3^{\text{rd}}$ continuum in the region around 200-350~nm, if assuming that the integral of the latter amounts to 0.0036~ph/eV, as recently measured in \cite{Santorelli}. These results are not presented here as they will be discussed in detail elsewhere. 

Indirectly, the low impact of recombination light in the window 210-800~nm for low CF$_4$ concentrations may be inferred from: i) its absence for mixtures above 1\% CF$_4$ (e.g., Fig. \ref{recombination_light} in appendix) for which the ionization densities are similar; ii) the fact that full charge-collection is reached to within less than 5\% for all conditions (e.g., Fig. \ref{CF4_1bar_currents}); iii) the fact that NBrS constitutes a featureless continuum above a certain wavelength threshold depending on the electron energy \cite{PRX} and, within that assumption, no significant field-induced modification of the characteristic UV and visible bands of the Ar/CF$_4$ scintillation could be observed below 1\% CF$_4$.

In the spectra shown in Fig. \ref{Spectra_0V_40mA} it can be seen that the transition between the pure-argon spectrum and the CF$_4$ one starts to happen as soon as 0.1$\%$ CF$_4$ is introduced. At that concentration, the appearance of a new peak at 290~nm and the small bump at around the CF$_3^*$ band hint towards a contribution beyond that of direct CF$_4$ excitation, that would be otherwise suppressed 1000 times relative to the 100\% CF$_4$ case. The $3^{\textnormal{rd}}$ continuum from argon quenches rapidly as CF$_4$ increases and it halves for just 0.2\% CF$_4$. For higher concentrations, the appearance of the CF$_4^{+,*}$ band associated to the $\tilde{C}\rightarrow\tilde{X}$ transition (centered at 230~nm) obscures the effect. As already noted, the near-infrared emission from argon displays self-quenching as the pressure increases, but it is also strongly suppressed in the presence of CF$_4$, becoming undetectable above 5\%CF$_4$ at 5~bar. The N$_2$ bands resulting from Ar$^*$ transfers disappear already at 0.1\% CF$_4$, suggesting that N$_2$ contamination is well below that concentration, if recalling that the Ar$^*$ quenching rates are comparable for the two molecules \cite{Setser}. The most prominent contamination in the system seems to be H$_2$O, that leads to OH$^*$ emission at around 310~nm and is arguably driven by Ar$^+$ + H$_2$O $\rightarrow$ Ar + H$_2$O$^+$ charge transfer \cite{OH_reco}. Given the shape-modification observed for the UV band of CF$_4$ at around 1~bar (where the presence of the OH$^*$ peak is most prominent relative to higher pressures), it cannot be fully excluded that H$_2$O might have a small influence in that case, specially for low CF$_4$ concentrations. For high pressures and high CF$_4$ concentrations, the OH$^*$ peak vanishes and the UV spectra stabilizes. Collisional-relaxation of CF$_4^{+,*}(v)$ sates down to the bottom of the potential well CF$_4^{+,*}(v=0)$ represents a plausible alternative, that would also explain the emergence of fully-formed UV bands when pressure and CF$_4$ concentration increases (as Ar is a priori inefficient for this process). 

\begin{figure}[h!]
\makebox[\linewidth]{\includegraphics[scale=0.18]{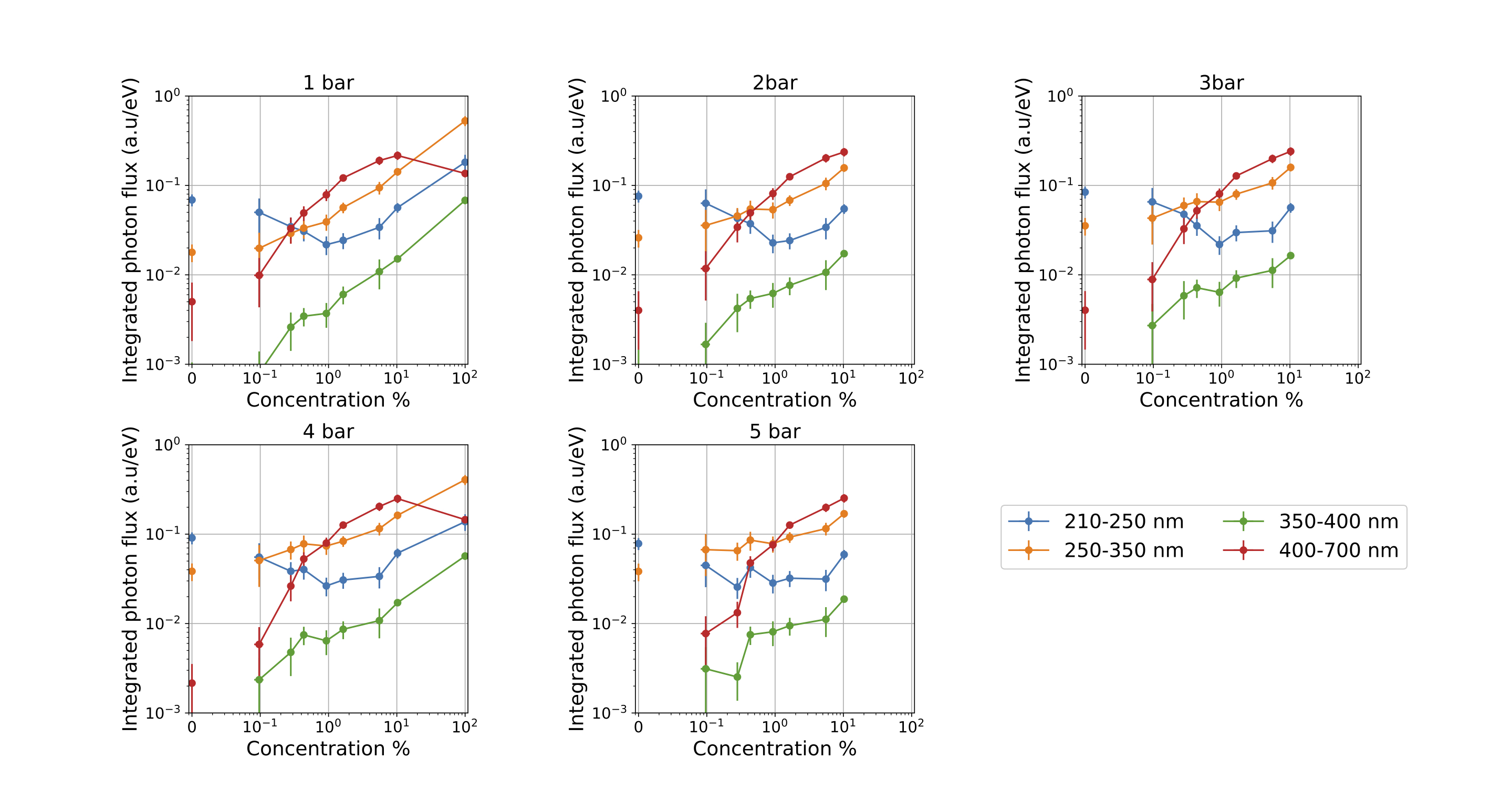}}
\caption{Integrated scintillation yields for the Ar/CF$_4$ system (per eV of released energy), shown in different bands as a function of CF$_4$ concentration, for different pressures. Zero-concentration yields have been added to the logarithmic $x$-axis to illustrate the asymptotic behaviour. (The argon peak located at 700~nm and the peaks caused by impurities were removed in this analysis) }
\label{Concentration_Yields_0V_40mA}
\end{figure}

Figure \ref{Concentration_Yields_0V_40mA} compiles the integrated yields in the most representative regions (210-250~nm, 250-350~nm, 350-400~nm and 400-700~nm) for different pressures and as a function of the CF$_4$ concentration. The trend of the 210-250~nm emission (blue) follows from the quenching of the Ar $3^{\textnormal{rd}}$ continuum, with CF$_4^{+,*}$ emission from $\tilde{C} \rightarrow \tilde{X}$ taking over as the CF$_4$ concentration increases, causing a minimum for concentrations around 1\% CF$_4$. In the other UV bands the increase is monotonous with CF$_4$ while the visible band shows an optimum for concentrations around or above the ones studied in this work. In general, visible-range yields are significantly increased over the ones in pure CF$_4$ in the range 2-10\%, and can be anticipated beyond the upper concentration studied in this work. A kinetic model addressing the observed behaviour is sketched in the next section.

\begin{figure}[h]
\makebox[\linewidth]{\includegraphics[scale=0.19]{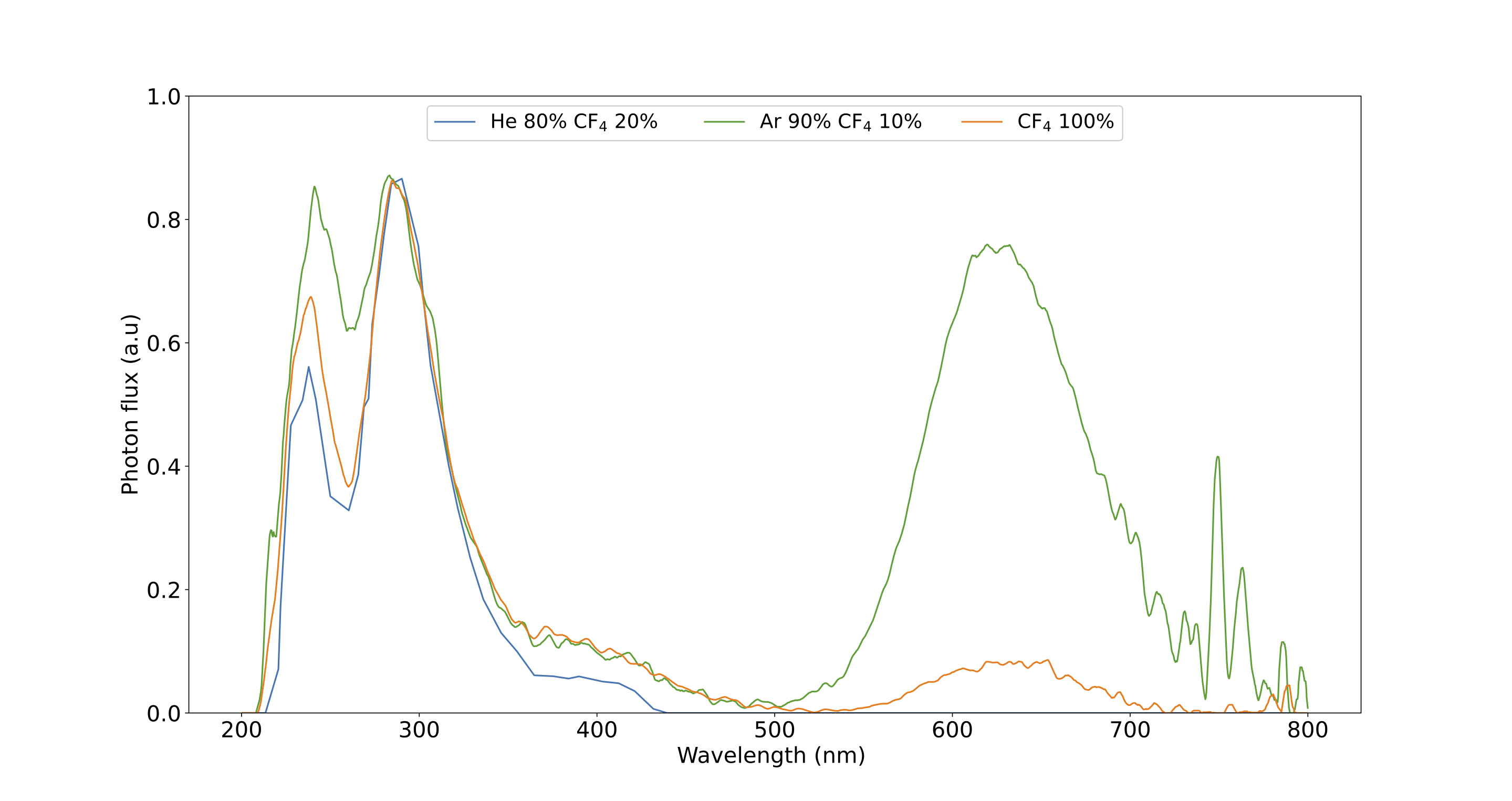}}
\caption{Comparison between the primary scintillation spectrum for pure CF$_4$ (orange), Ar/CF$_4$ at 10\% per volume (green) and He/CF$_4$ at 20\% per volume (blue), at 1~bar. All spectra have been arbitrarily normalized to the 290~nm UV peak.}
\label{Spectra_WLS}
\end{figure}

\section{Discussion} \label{Kinetic}

\subsection{Wavelength-shifting pathways}
Our data presents strong evidence of wavelength-shifting and in particular Fig. \ref{Concentration_Yields_0V_40mA} suggests, qualitatively, that scintillation in the 630~nm band (CF$_3^*$) must feed from Ar$^{*}$ transfers and not just through direct CF$_3^*$ formation, otherwise a proportional trend would be expected. The observed increase (approximately proportional up to 1\%CF$_4$) shows a drop above or around 10\%CF$_4$ and might still be attributed to direct CF$_3^*$ formation followed by self-quenching with CF$_4$. However this would imply a strong dependence with pressure, that is not seen in data. Fig. \ref{Spectra_WLS} shows for illustration a spectral comparison with the CF$_4$ and He/CF$_4$ systems, for which the CF$_3^*$ band appears depopulated relative to the UV one, adding further support to the role of Ar$^{*}$ states at CF$_3^*$ formation. Given that the threshold for CF$_3$ production sits at 12.5 eV \cite{Winters} 
and the one for CF$_3^+$ production at around 16 eV \cite{Christophorou, Zhang}, the threshold for CF$_3^*$ production must lie in between. This indicates that the Ar$^{*}$ state(s) involved in transfers lie well above the lowest-lying Ar excited states at 11.5~eV and close to the continuum (IP$_{Ar}$=15.7~eV), thereby labeled Ar$^{**}$ hereafter. A discussion on the nature of such a state(s) is postponed to the end of this section. On the other hand, the UV scintillation may be tentatively attributed to transfers involving the higher-lying $3^{\textnormal{rd}}$ continuum precursors (Ar$^{*,+}_{n=2,3}$, Ar$^{++}_{n=2,3}$) \cite{Wieser}. The fact that the yields in the region 210-250~nm and 250-350~nm show opposing trends up to around 1\% CF$_4$, with the total yield remaining approximately constant, is a good indicator that the energy is being transferred between species. The extracted quenching rates and spectral shapes add further support to this interpretation, as shown later. 

\begin{figure}[h]
\makebox[\linewidth]{\includegraphics[scale=0.45]{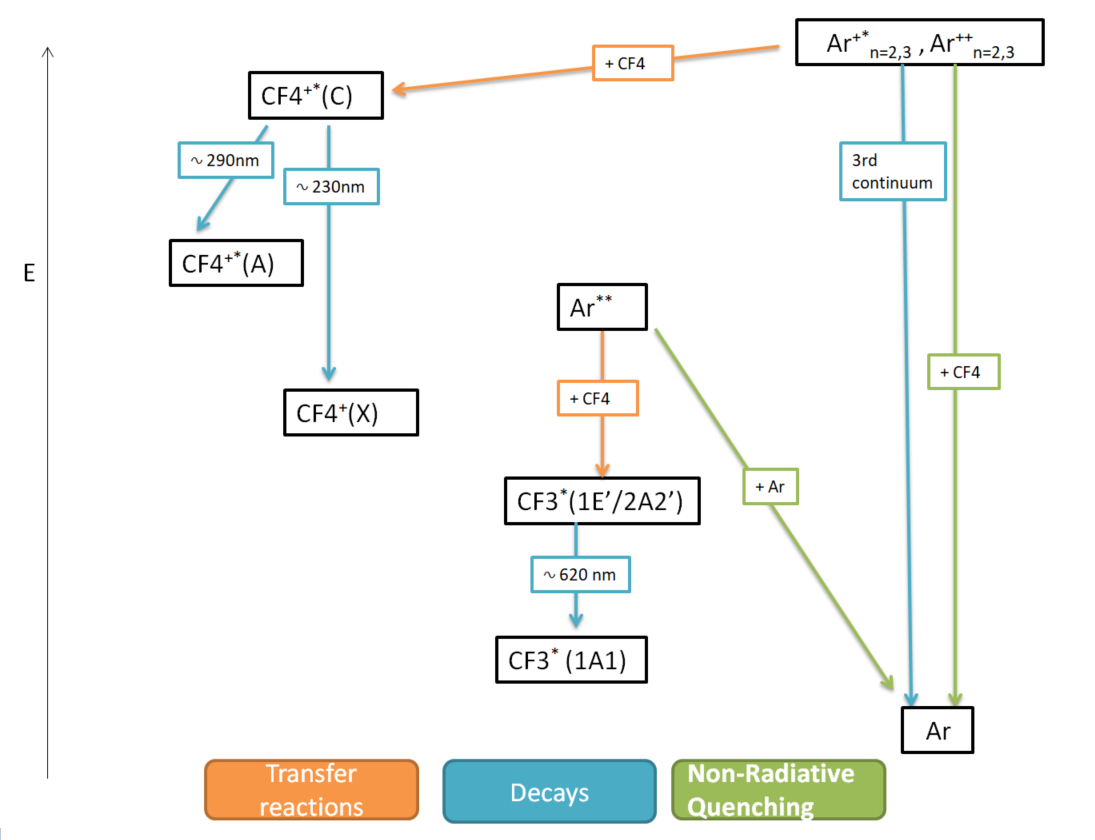}}
\caption{Kinetic scheme use to describe wavelength-shifting in Ar/CF$_4$ mixtures. The main scintillation drivers are the CF$_4^{+,*}$, CF$_3^*$ and Ar 3$^{\textnormal{rd}}$ continuum states. Energy considerations lead to the hypothesis of an additional high-lying Ar$^{**}$ state, whose nature is discussed in text.}
\label{Model_scheme}
\end{figure}

A kinetic model has been developed keeping the above considerations in mind, in order to quantitatively interpret our experimental results. It is sketched in Fig \ref{Model_scheme} and detailed in the following. 


Aiming at a reduced number of model parameters, the $3^{\textnormal{rd}}$ continuum precursors are characterized through an effective decay constant of 5~ns (e.g., \cite{Santorelli}):
\begin{eqnarray}
    \ce{&Ar$^{+,*}_{n=2,3}$, Ar$^{++}_{n=2,3}$ &->[$\tau_{3^{rd}}$] Ar$^+$ + Ar$^{(+)}$ (  + Ar) + h$\nu$ (180 - 300 nm)} 
\end{eqnarray}
For CF$_4$, the states that leave a clear footprint in the spectra can be matched to the following decays:

\begin{eqnarray}
    \ce{&CF$_3^*(2A_2^{''}, 1E^{'})$ &->[$\tau_{CF_3}$] CF$_3(1A_1^{'})$ + h$\nu$  (630 nm) } \label{CF4_rad1}\\
    \ce{&CF$_4^{+,*}(\tilde{C})$ &->[$\tau_{CF_4(C->X)}$] CF$_4^+(\tilde{X})$
    + h$\nu$  (230 nm)}\\\label{CF4_rad2}
    \ce{&CF$_4^{+,*}(\tilde{C})$ &->[$\tau_{CF_4(C->A)}$] CF$_4^{+,*}(\tilde{A})$
    + h$\nu$  (290 nm)} \label{CF4_rad3}
\end{eqnarray}  
To the best of our knowledge, not all the possible decays of the accessible CF$_4^{+,*}$ states have been observed in literature, e.g. \ce{CF$_4^{+,*}$(B) -> CF$_4^{+,*}$(A)}. As CF$_4^{+,*}$ states are unstable \cite{Zhang}, it is possible that dissociation out-competes radiative decay in some of them. For the states involved in eqs. \ref{CF4_rad1}-\ref{CF4_rad3}, however, we opted to assign a decay probability of 100\% to avoid the introduction of new additional parameters. As shown later, a good $\chi^2$ is obtained in the proposed kinetic scheme through a global fit employing 2 parameters per spectral region (UV and visible), so a further increase in the number of parameters was deemed unnecessary.

  

Transfer reactions between Ar states and CF$_4$ represent the last ingredient. For the visible component, we make the natural assumption that transfers between the Ar$^{**}$ state(s) and CF$_4$ compete just with self-quenching, summarized as:

\begin{eqnarray}
\label{eq:ar_transf}
    \ce{&Ar$^{**}$ + CF$_4$ 
    &->[$k_{Ar^{**}->CF_3^*}$] Ar + CF$^*_3(2A_2^{''}, 1E^{'})$ }
    \\
\label{eq:ar_quench}
    \ce{&Ar$^{**}$ + Ar 
    & ->[$k_{Ar^{**}-> Ar^*}$] Ar$^{*}$ + Ar} 
\end{eqnarray}
It is in principle possible to include an additional non-radiative quenching channel of Ar$^{**}$ with CF$_4$, that has been neglected again on the basis that it is not needed to describe data and it would add unnecessary complexity to the model. Within the proposed kinetic scheme, Ar$^{**}$ transfers would lead to CF$^*_3$ scintillation with near-100\% probability.

Last, we consider transfer reactions leading to UV emission between the $3^{\textnormal{rd}}$ continuum precursors and CF$_4$, together with a quenching reaction to non-radiative states:
\begin{eqnarray}
\label{eq:3rd_transf}
    \ce{&Ar$^{+,*}_{n=2,3}$ (Ar$^{++}_{n=2,3}$) + CF$_4$  
    &->[$k_{Ar^{3rd}->CF_4^{+,*}}$] Ar + CF$_4^{+*}(\Tilde{C})$ }\\
\label{eq:3rd_quench}
    \ce{&Ar$^{+,*}_{n=2,3}$ (Ar$^{++}_{n=2,3}$) + CF$_4$  &->[$k_{Ar^{3rd}->Ar}$] \textnormal{non-radiative}}
\end{eqnarray}
where rates for transfer and non-radiative quenching have been introduced for an `effective' $3^{\textnormal{rd}}$ continuum precursor. Along this line, reactions of $3^{\textnormal{rd}}$ continuum states with ground-state Ar are assumed to be already accounted for when considering the kinetics of such an `effective' precursor, and remain unaltered in the presence of CF$_4$ (for a detailed pathway scheme of the $3^{\textnormal{rd}}$ continuum formation, the reader is referred to \cite{Wieser}). In the following, reaction rates $[t^{-1}]$ are defined for 1~bar of the reactive, and scaled based on pressure and species concentration (as done for instance in \cite{Micro_Diego}).

    
    

Before evaluating the model, it should be noted that the possibility of  self-quenching of the CF$_3^*(2A_2^{''}, 1E^{'})$ state with CF$_4$ has been omitted due to the negligible pressure-dependence of Ar/CF$_4$ scintillation in the 630~nm band. CF$_4^{+,*}$ states, on the other hand, evidence a small self-quenching on the UV region in pure CF$_4$, compounded with the aforementioned indications of collisional relaxation for Ar/CF$_4$ mixtures at low pressures and CF$_4$ concentrations. Such dependences with pressure can be easily included in the model but, being a small effect and not shedding light into the main transfer mechanisms, they have been omitted for the sake of simplicity. The near-visible band centered around 364 nm, arising from the $\tilde{D} \rightarrow \tilde{C}$ transition, is also not considered given its relatively small contribution to the total spectrum.

From the above set of reactions, it is possible to derive the scintillation probability (per eV of energy deposited in the medium) 
of the states CF$_3(2A_2^{''}, 1E^{'})$, CF$_4^{+,*}(\Tilde{C})$, and of the effective  state Ar$^{\textnormal{3rd}}$:


\begin{equation}
\begin{split}
\label{N_ph_cf3}
 P_{\gamma, CF_3^*} = & f_{CF_4} \cdot P_{\gamma,CF_3^*}\big|_{dir}+
 \left((1-f_{CF_4}) \cdot P_{Ar^{**}} \cdot \frac{K_{Ar^{**}->CF_3^{*}}}
 {K_{Ar^{**}->CF_3^{*}}
 + \frac{(1-f_{CF_4})}{f_{CF_4}} \cdot K_{Ar^{**}->Ar^*}}\right)\\
\end{split}
\end{equation}


\begin{equation}
\label{N_ph_cf4}
\begin{split}
 P_{\gamma, CF_4^{+*}} = &f_{CF_4} \!\cdot\! P_{\gamma,CF_4^{+*}}\big|_{dir}+\!  
 \left(\!(1\!\!-\!\!f_{CF_4}) \!\cdot \!P_{Ar^{3rd}} \!\cdot \!\frac{f_{CF_4} \! \cdot \! n \! \cdot \! K_{Ar^{3rd}->CF_4^{+*}}}
 {1/\tau_{3rd} + f_{CF_4}\! \cdot \!n \!\cdot \!( K_{Ar^{3rd}->CF_4^{+*}} \!+ \!K_{Ar^{3rd}->Ar})}\!\right)\\
\end{split}
\end{equation} 

\begin{equation}
\begin{split}
\label{N_ph_3rd}
 P_{\gamma,Ar^{\textnormal{3rd}}}= (1-f_{CF_4}) \cdot P_{Ar^{3rd}} \cdot
 \left( \frac{1/\tau_{3rd}}{1/\tau_{3rd} + f_{CF_4} \cdot n \cdot (K_{Ar^{3rd}->CF_4^{+*}}
 +K_{Ar^{3rd}->Ar})}\right)
 \end{split}
\end{equation}
Here $f_{CF_4}$ represents the CF$_4$ concentration, $n$ equals the pressure ratio $P/P_0$, and the quenching and transfer rates are the ones defined in \ref{eq:ar_transf}-\ref{eq:3rd_quench}. Parameters with subscript `dir' refer to the probability of direct scintillation, and $P_{Ar^{**}}$, $P_{Ar^{\textnormal{3rd}}}$ stand for the formation probability of the Ar$^{**}$ state(s) and of the $3^{\textnormal{rd}}$ continuum precursors, respectively. Some of the parameters needed to evaluate the above equations can be constrained based on existing experimental data, including this work. The direct scintillation probabilities, for instance, as well as the probability of formation of $3^{\textnormal{rd}}$ continuum precursors, can be obtained from the yields in pure CF$_4$ and pure argon. 
The time constant for the argon $3^{\textnormal{rd}}$ continuum has been taken from \cite{Santorelli}. Further, based on the model structure, only the ratio of Ar$^{**}$ transfer to quenching rates in \ref{N_ph_cf3} is relevant, reducing the total number of fit parameters to four. The other three parameters represent the formation probability of Ar$^{**}$ states, and the non-radiative quenching and transfer rates of CF$_4^{+,*}$. The model structure also makes explicit the lack of pressure-dependence of CF$_3^*$ scintillation observed for any admixture.

\begin{figure}[h]
\makebox[\linewidth]{\includegraphics[scale=0.19]{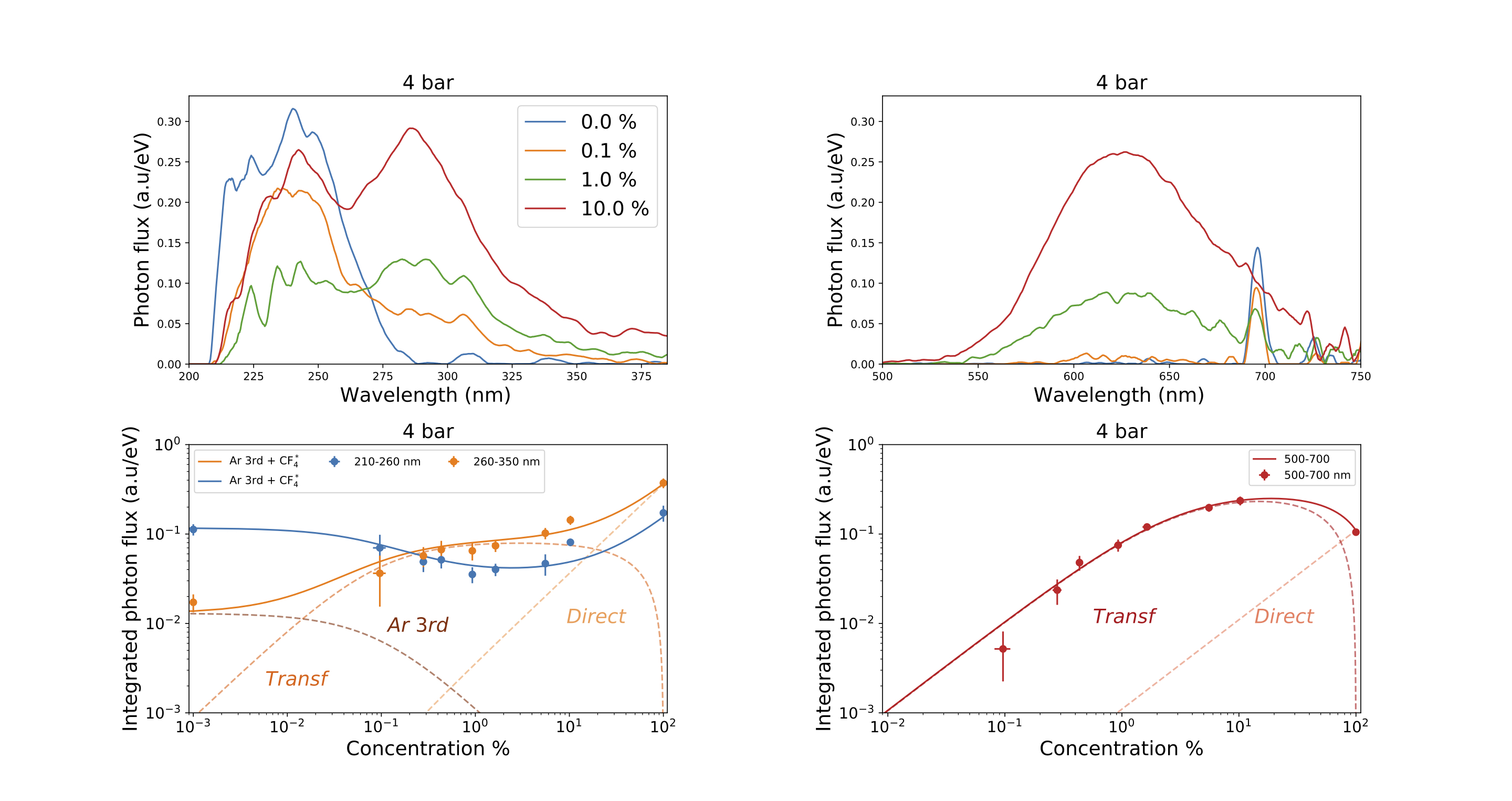}}
\caption{Top row: scintillation spectra for different CF$_4$ concentrations at 4~bar, zoomed in the ultraviolet (left) and visible (right) regions. Bottom row: integrated yields on the ultraviolet (left) and visible (right) regions (shown as closed circles), superimposed to the kinetic model introduced in text. The experimental value for zero concentration  is added on the bottom-left plot at 0.001\% CF$_4$, taking advantage of the fact that the model asymptotically tends to a constant in that case.}
\label{Model}
\end{figure}

A weighted global fit of the proposed kinetic model to the three data series associated with the 230, 290 and 630~nm bands was performed. Yields in the first two bands were fitted to a sum of eqs. \ref{N_ph_cf4} and \ref{N_ph_3rd} and the 630~nm band was described through eq. \ref{N_ph_cf3}. The fit is shown in Fig \ref{Model} for a pressure of 4~bar (blue, orange, red lines), alongside the corresponding experimental data (full circles). 
Its reduced $\chi^2$ of 1.56 adds plausibility to the present interpretation. The following values and uncertainties were obtained for the fit parameters: P$_{Ar^{**}}/P_{Ar^{3rd}} = 3.19 \pm 0.39$ (population of Ar$^{**}$ relative to that of 3$^{\textnormal{rd}}$ continuum precursors); $\frac{K_{Ar^{**}->CF_3^{*}}}{K_{Ar^{**}->Ar^*}} = 36.5 \pm 7.9$ (ratio of transfer to collisional quenching of Ar$^{**}$); K$_{Ar^{3rd}->CF_4^{+*}} = 49 \pm 18 \quad $ns$^{-1}$ (transfer rate of the Ar $3^{\textnormal{rd}}$ continuum) and K$_{Ar^{3rd}->Ar} = 4.1 \pm 3.3 \quad $ns$^{-1}$ (non-radiative quenching of the Ar $3^{\textnormal{rd}}$ continuum). Within the proposed kinetic model, the ratio of transfer-mediated scintillation to direct scintillation can be computed: for 1\% CF$_4$, illustratively, values as large as $72 \pm 20 $ (visible) and  $20.9 \pm 2.1$ (UV) are obtained.


In the UV-band, a detailed analysis of the spectral shapes (Fig. \ref{Model} top-left) brings additional support to the proposed interpretation: as soon as 0.1\% CF$_4$ is added to argon, there is a strong suppression of the argon $3^{\textnormal{rd}}$ continuum, coinciding with the appearance of the 290~nm peak from CF$^{+,*}_4$. For 1\% CF$_4$, the effect is even more clear. Taking as a reference the yields measured earlier for $\alpha$-particles in the Ar $3^{\textnormal{rd}}$ continuum (around 3500~ph/MeV \cite{Santorelli}) and in the VUV-visible range of CF$_4$ (1000-3000~ph/MeV \cite{Pansky, CF4_alphas_Moro, Rumore, Lehault}), it is implausible that direct CF$_4$ excitation could be responsible for the observed levels of X-ray scintillation when the species is present at a mere sub-percent level. Above 1\% CF$_4$, the overlap between argon $3^{\textnormal{rd}}$ continuum and the CF$_4^{+,*}$ emission at 230~nm complicates a qualitative description, and a direct comparison with the proposed kinetic model has to be used instead (Fig. \ref{Model} bottom-left).
In any case, the fact that the number of photons emitted in the UV region is maintained approximately constant when transfer reactions dominate (i.e., below a few \% CF$_4$) reinforces the idea that the kinetics of the Ar $3^{\textnormal{rd}}$ continuum states drives the UV scintillation of the admixture, and that the CF$^{+,*}_4$ state scintillates efficiently upon transfer, as assumed. Although the agreement seems convincing, additional support to the proposed pathway scheme can be found: according to the model, the scintillation of the argon $3^{\textnormal{rd}}$ continuum gets quenched down to just $8.6\% \pm 2.8\%$ of its nominal value in the pure gas when in presence of 1\% CF$_4$. This is compatible with the quenching level that has been reported for the Ar $3^{\textnormal{rd}}$ continuum when admixed with CO$_2$, at about 1\% concentration \cite{Strickler}. Given that both CF$_4$ and CO$_2$ are energetically accessible to transfers from the high-lying Ar$^{+,*}_{2,3}$ and Ar$^{++}_{2,3}$ states, and the similar molecule size, this approximate agreement is reassuring.

In the visible band, the proposed model where CF$^*_3$ formation competes with self-quenching provides a natural explanation for the total absence of pressure dependencies. It requires, however, the somewhat artificial introduction of a new Ar$^{**}$ state or set of states. Invoking the 3$^{\textnormal{rd}}$ continuum precursors cannot be excluded, although it would require increasing the model complexity substantially: the behaviour of the transfer reactions involved in UV and visible scintillation (Fig. \ref{Model}-bottom) is too different to be easily attributable to the same state. As discussed earlier, the existence of Ar$^{**}$ states is justified by energy considerations, given that they must be several eV above the Ar$^*$ $2^{\textnormal{nd}}$ continuum precursors at around 11.5~eV, yet below the argon IP (15.7~eV) in order to induce excited dissociation of CF$_3^*$. According to our model, the formation probability of Ar$^{**}$ is about $\times 3$ that of Ar$^{\textnormal{3rd}}$ states, that would imply (if using for reference the $2^{\textnormal{nd}}$ and $3^{\textnormal{rd}}$ continuum yields measured in \cite{Santorelli}) a substantial part of the available Ar excited states being eligible for transfer (about 50\%). Also, the transfer rate would need to be nearly 40 times larger than self-quenching with Ar ($36.5 \pm 7.9$ from our fit) and it remains open which mechanism could cause such large transfer values. Although the role of the Ar$^{**}$ states cannot be excluded from present results, there is an alternative explanation in the formation of an ArCF$_3^*$ exciplex. There is apparently no information in literature about this process that is, e.g., generally absent in the modelling of Ar/CF$_4$ discharges \cite{DIS1, DIS2, DIS3}. ArCF$_3$ is iso-electronic with CF$_4^-$ that is in fact known not to be stable, except for (CF$_4)_n$ clusters \cite{CF4-}. There are suggestions of the formation of the (expectedly more stable) CF$_4^{-,*}$ state in some works (e.g. \cite{CF4-*}), but not enough evidence is provided. If the binding energy of such an exciplex would be on the order of few eV, it is conceivable that it could be formed starting from the long-lived Ar$_2^*$ triplet state. Although speculative at the moment, such a mechanism would provide a natural explanation for the lack of pressure-dependence, the similarity between the populations of Ar$^{**}$ and $2^{\textnormal{nd}}$ continuum precursors, as well as for the preponderance of wls-transfers in Ar compared to, e.g., He.

\subsection{Comparison with previous results}

The results obtained in this work may be compared with the ones obtained for Ar/CF$_4$ mixtures in a 9~MeV proton beam at 1~bar in \cite{Liu}. Little details are found there in regard to space charge, recombination and beam-induced scintillation and in fact the relative normalization of the visible/UV bands is not given either. Qualitatively, it is possible to see that the relative increase in both the visible and UV bands from 1\% to 10\% CF$_4$ concentrations is around a factor 3, compatible with present results. The region above 700~nm is characterized by the presence of an additional molecular emission while the Ar IR emission appears fully quenched, both observations being in stark contrast with our results. The absence of data for pure CF$_4$ together with the strong contamination found for argon data in that work, preclude any estimate of the photon yields or wavelength-shifting capability. 

A spectral comparison between scintillation induced by X-rays (this work) and $\alpha$-particles (in \cite{CF4_alphas_Moro}) seems more reliable at this point, and can be seen in Fig. \ref{Pure_gases_spectra} (orange, dashed). Both spectra were obtained at 1~bar, arbitrarily normalized to the 290~nm peak. They display an approximate agreement in the UV and blue regions, however the emission in the red region appears off by a factor of 2.8. The discrepancy is preserved when considering the ratio between the other two UV peaks and the CF$_3^*(2A_2^{''}, 1E^{'})\rightarrow$ CF$_3 (1A_1^{'})$ one at 630~nm. These transitions are well above the calibration mark of the lamp at 300~nm and also far enough into the visible region so that we can safely exclude any strong wavelength-asymmetry of the light collection process in the chamber compared to the calibration setup. As no hints of charge recombination were observed at 1~bar neither in \cite{CF4_alphas_Moro} nor in this work, these measurements point to a fundamental difference between the scintillation mechanisms for $\alpha$ particles and X-rays in CF$_4$.

Last, it must be recalled that the strength of CF$_4$ scintillation in the VUV-visible range, as obtained for $\alpha$ particles at 1~bar, is currently found at levels of $1000$-$3000$~ph/MeV \cite{Pansky, CF4_alphas_Moro, Rumore, Lehault}. Values within this range have been reported, too, for Ar/CF$_4$ mixtures around 10~bar in \cite{We1}. Nonetheless, the large experimental spread on the above CF$_4$ yields, together with the particle-dependence of the spectral emission reported here, call for future studies on the scintillation yields of these type of mixtures.

\section{Conclusions} \label{conc}

We have presented a comprehensive data-set on the primary scintillation spectra of Ar/CF$_4$ mixtures in the pressure range 1-5~bar and CF$_4$ concentrations from 0.1 to 10\%, including pure gases. Our results, obtained under strong X-ray irradiation yet in conditions shown to be free from recombination and space charge effects, provide a clear indication that Ar $3^{\textnormal{rd}}$-continuum precursors play a pivotal role in the UV-scintillation of Ar/CF$_4$ mixtures. On the other hand, a high-laying Ar$^{**}$ state or an ArCF$_3^*$ exciplex seem the most plausible candidates leading to CF$_3^*$ formation (responsible for the scintillation in the visible range), with a simple pathway scheme explaining the observed phenomenology, in particular the lack of pressure-dependence of the measured yields. The proposed kinetic model resorts to just 4 parameters (2 per emission band), achieving a satisfactory agreement with a reduced $\chi^2$ of 1.56. A more complex mechanism starting from the precursors of the Ar $3^{\textnormal{rd}}$ continuum could still be advocated to cause scintillation in the visible range, however its elucidation does not seem accessible to present experimental conditions.

In sum, wavelength-shifting in the Ar/CF$_4$ system is very strong for the conditions studied: at a mere 2\% CF$_4$, for instance, scintillation in the 500-700~nm (CF$_3^*$) band exceeds that of pure CF$_4$ with independence from pressure. UV scintillation remains at strengths comparable to the visible one in the concentration range 1-10\% CF$_4$, and progressively dominates outside it. Overall, upon just 1\% CF$_4$ addition, the ratio of transfer-mediated scintillation to direct scintillation is estimated to be as large as $72 \pm 20 $ (visible) and $20.9 \pm 2.1$ (UV). 

Our measurements convey as well strong evidence of the dependence of the spectra of emission on particle type. The ratio of the UV/visible bands, as observed for X-rays in this work, is about $\times 2.8$ larger than measured earlier for $\alpha$'s in pure CF$_4$ at around 1~bar, both performed in recombination-free conditions. Overall, the presented results show great promise for technological applications in future particle detectors in the fields of rare event searches, nuclear and neutrino physics.

\section*{Acknowledgements}

The authors want to thank Saulo Vázquez (USC) for his valuable insights on the chemistry of Ar/CF$_4$ reactions and L. Margato (LIP-Coimbra) for useful feedback. This research was funded by the Spanish Ministry (`Proyectos de Generación de Conocimiento', PID2021-125028OB-C21), Xunta de Galicia (Centro singular de investigación de Galicia, accreditation 2019-2022), and by the “María de Maeztu” Units of Excellence program MDM2016-0692. DGD was supported by the Ramón y Cajal program (Spain) under contract number RYC-2015-18820.

\appendix

\section{Electric field in the ionization volume} \label{EF_section}

\begin{figure} [h!!!]
\includegraphics[width=.6\textwidth]{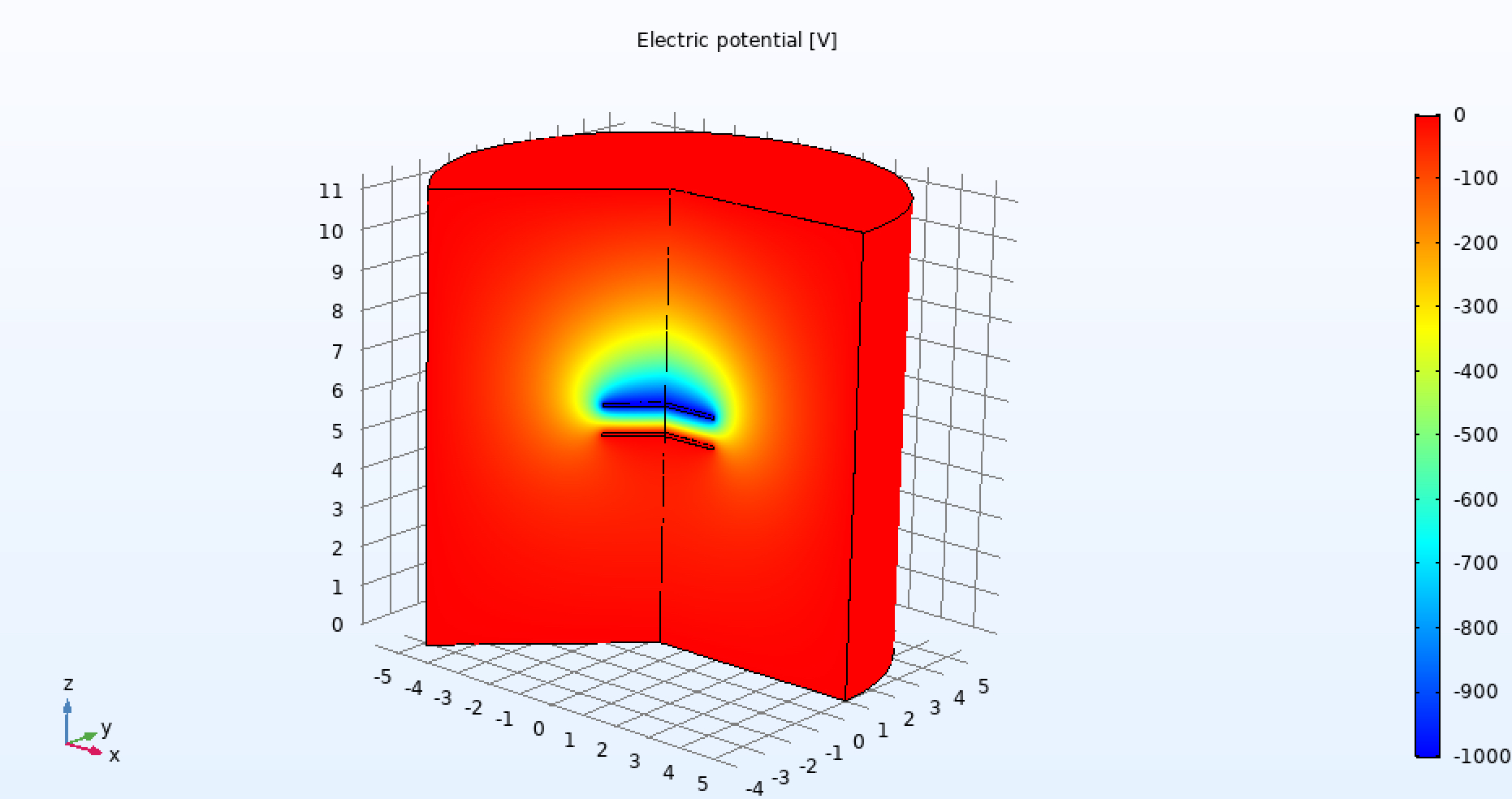}\hfill
\includegraphics[width=.4\textwidth]{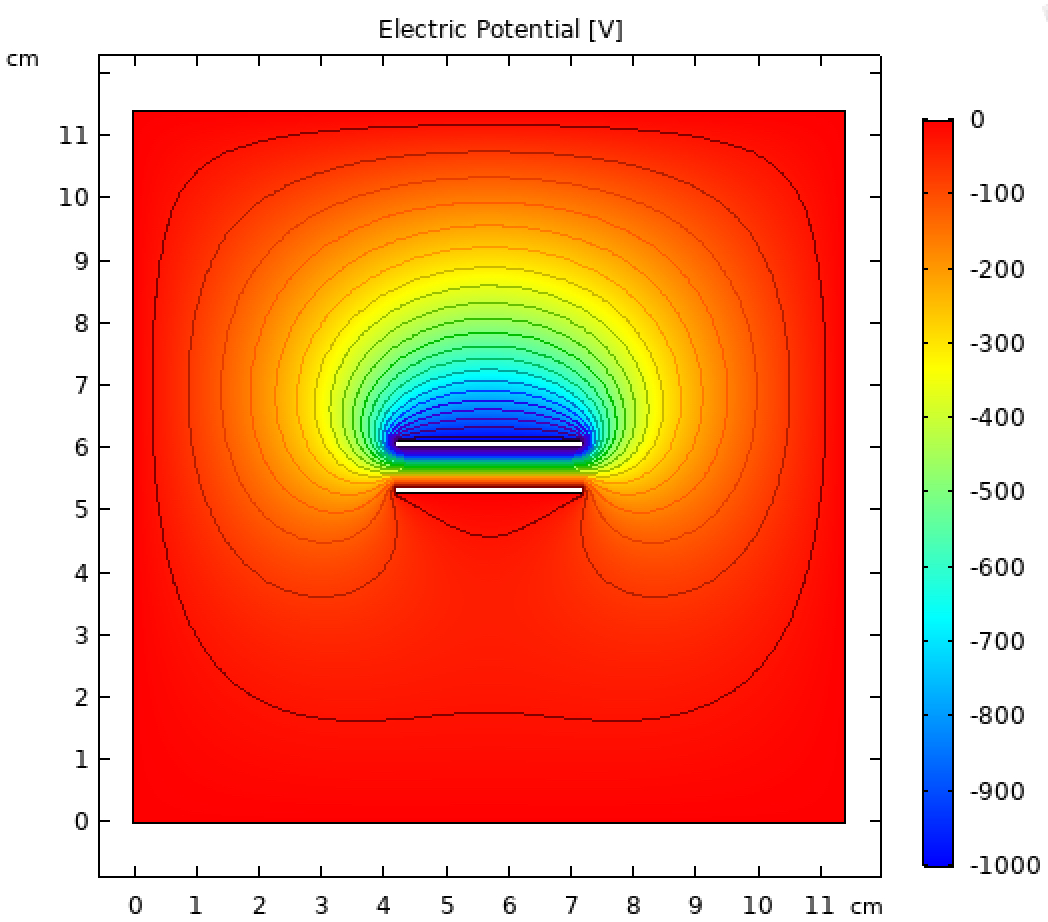}\hfill
\includegraphics[width=1\textwidth]{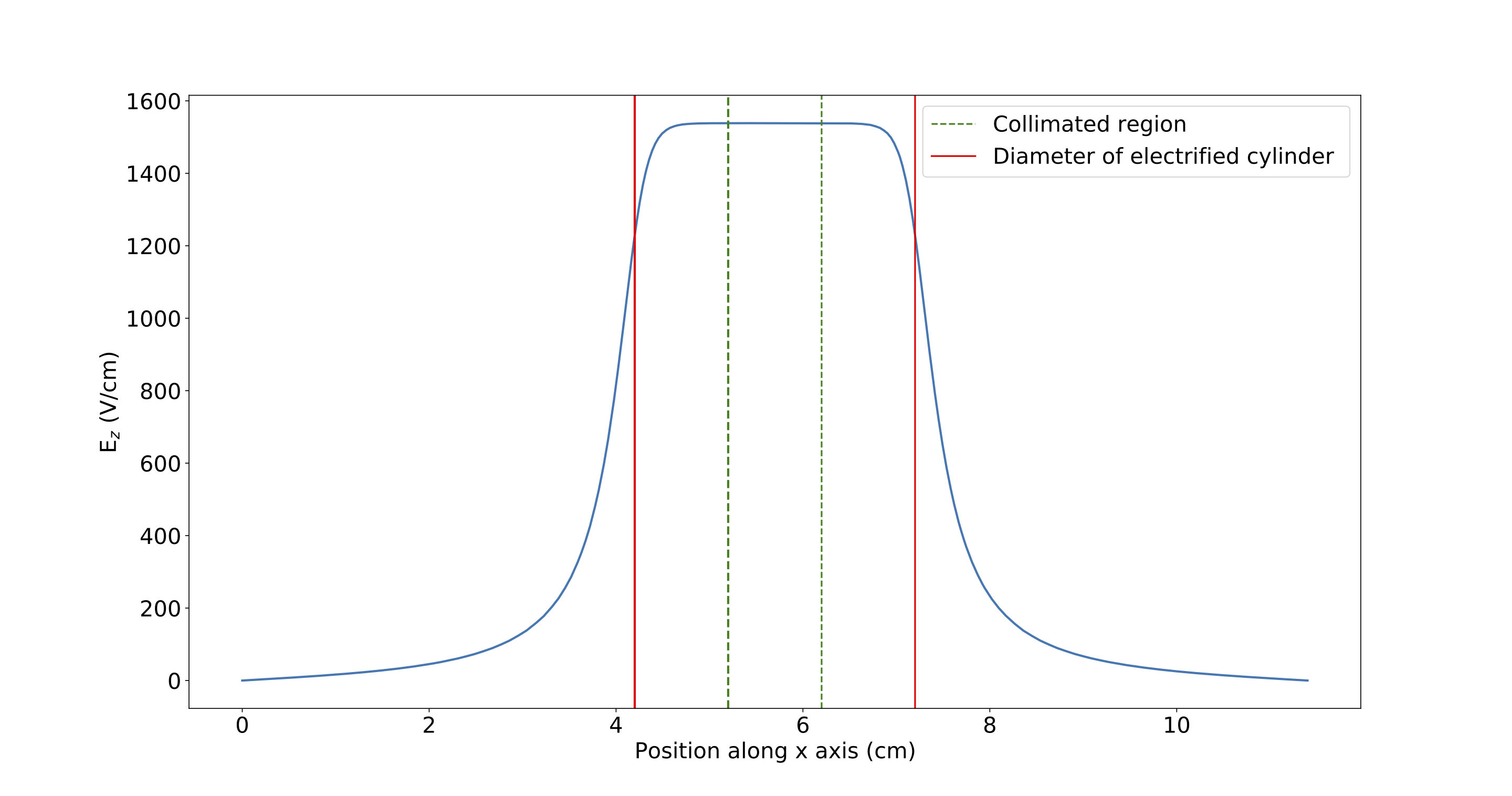}\hfill
\caption{Results from an axisymmetric COMSOL simulation, with the X-ray beam impinging from the positive $z$-axis. Top-left: geometry model and electric potential in 3D. Top-right: cross-section at mid-chamber and equipotential curves. Bottom: electric field component perpendicular to the electrodes at mid-chamber (symmetry plane of the cylinder), as a function of the distance from one of the chamber walls.}
\label{Chamber_field}
\end{figure}

An axisymmetric finite-element simulation of the electrified cell and the vessel walls was performed with COMSOL Multiphysics\textsuperscript{\textregistered} \cite{COMSOL}. As shown in Fig. \ref{Chamber_field}-bottom, the electric field is uniform over a region slightly exceeding 2~cm, considerably larger than the size of the ionization volume (collimated down to 1~cm at the chamber entrance -green dashed lines). The values of the electrical potential in 3D, together with the equipotential curves, are shown in Fig. \ref{Chamber_field}-top.

Space charge effects stemming from the presence of slowly-drifting ions can modify the electric field, leading to charge losses due to enhanced charge recombination, or to imperfect electron collection at the anode.
For the conditions of our measurements, where the ionization is highly uniform and the external electric field has been established to be uniform within the ionization region, the calculation performed in \cite{Space_charge_Mcdonald} can be used to assess the situation. In this case, the impact of space charge may be characterized with a single dimensionless parameter $\alpha$:

\begin{equation}
\alpha = \frac{D}{E_0} \sqrt{\frac{K}{\epsilon\mu}}
\label{alpha_equation}
\end{equation}
Here $D$ is the conversion gap (0.75~cm), $E_0 = V/D$ is the nominal electric field in the absence of space charge, $V$ the voltage drop between anode and cathode, $\epsilon\simeq\epsilon_0$ the electric permittivity of the medium, $\mu$ the positive-ion mobility and $K$ the rate of creation of electron-ion pairs per unit volume. The latter (in units of [$\mu$C/s/cm$^3$]) was determined at high electric fields, in conditions of full charge-collection.

\begin{figure}[h!!!]
\centering 
\hspace*{-1cm}\includegraphics[width=1.2\textwidth,origin=c,angle=0]{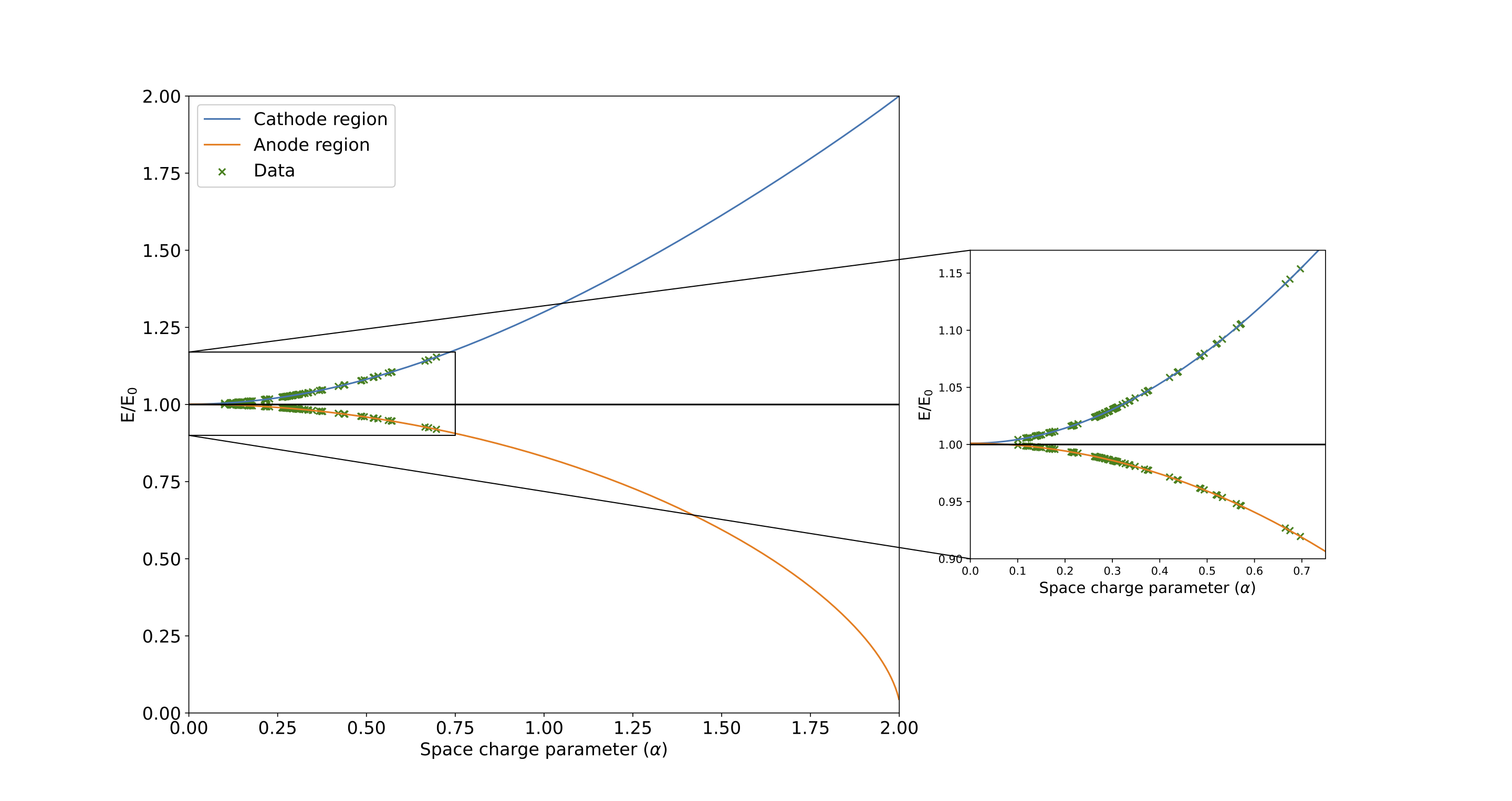}
\caption{Electric field distortion near cathode/anode as a function of the $\alpha$ parameter (blue/orange lines). Pairs of experimental values obtained in conditions of full charge-collection are overlaid, for all pressures, mixtures and tube intensities employed in this work (crosses).}
\label{alphas}
\end{figure}

The input mobilities were obtained in the following way: in argon, the main drifting ions were assumed to be Ar$^+$ and their mobility taken from \cite{RolandiBook}. In the case of Ar/CF$_4$ mixtures and pure CF$_4$, CF$_3^+$ ions were considered instead and their mobility taken from direct measurements in \cite{Santos}. CF$_3^+$ ions are produced as a byproduct of dissociation/decay of CF$_4^{+,*}$ states \cite{Zhang}, or through charge transfer reactions with CF$_4$ \cite{Santos}:

\begin{equation}
\ce{Ar$^{+}$ + CF$_4$ ->[$k_0$] CF$_3^+$ + Ar + F}
  \end{equation}
The electric field has a numerical dependence on $\alpha$, that can be computed following \cite{Space_charge_Mcdonald}. As the maximum variations take place close to the anode and cathode, we chose these two cases for representation in Fig. \ref{alphas}, with the experimental conditions corresponding to full charge-collection overlaid (crosses). Based on these results, the field distortions associated to space charge effects are shown to be minimal: values are typically at the $\sim$5$\%$-level or below, with a maximum field distortion of $\sim$15$\%$.

\section{Charge recombination} \label{section:Charge_recombination}

Electron-ion and ion-ion recombination, either geminal \cite{ONSAG}, columnar \cite{JAFFE} or volume \cite{VolumeReco} can reduce the collected charge, so it is important to exclude these during measurements. On general grounds, given the relatively low pressures and X-ray energies, the main contribution to charge recombination is expected to come from the volume, and so being associated to the setup geometry and not to the particle type. Therefore any charge recombination present in the system would render the interpretation of the results problematic and must be avoided. During the measurements, the electric field was increased, for any given gas mixture and pressure, up to the point where full charge-collection could be guaranteed, within less than 5\% variation relative to the flat-top (see Fig. \ref{CF4_1bar_currents}-left for the case of pure CF$_4$). This already assures that all charge is being collected, except perhaps for stray electrons from diffusion, fringe fields or space charge, effects that are expected to be below 5-10\% level as per the discussions in previous sections. The good correlation between ionization rate and tube intensity across all the ionization densities considered (Fig. \ref{CF4_1bar_currents}-right) confirms the negligible presence of charge recombination. 


\begin{figure}[h!!!]
\centering
\hspace*{-1.5cm}\includegraphics[width=1.2\textwidth]{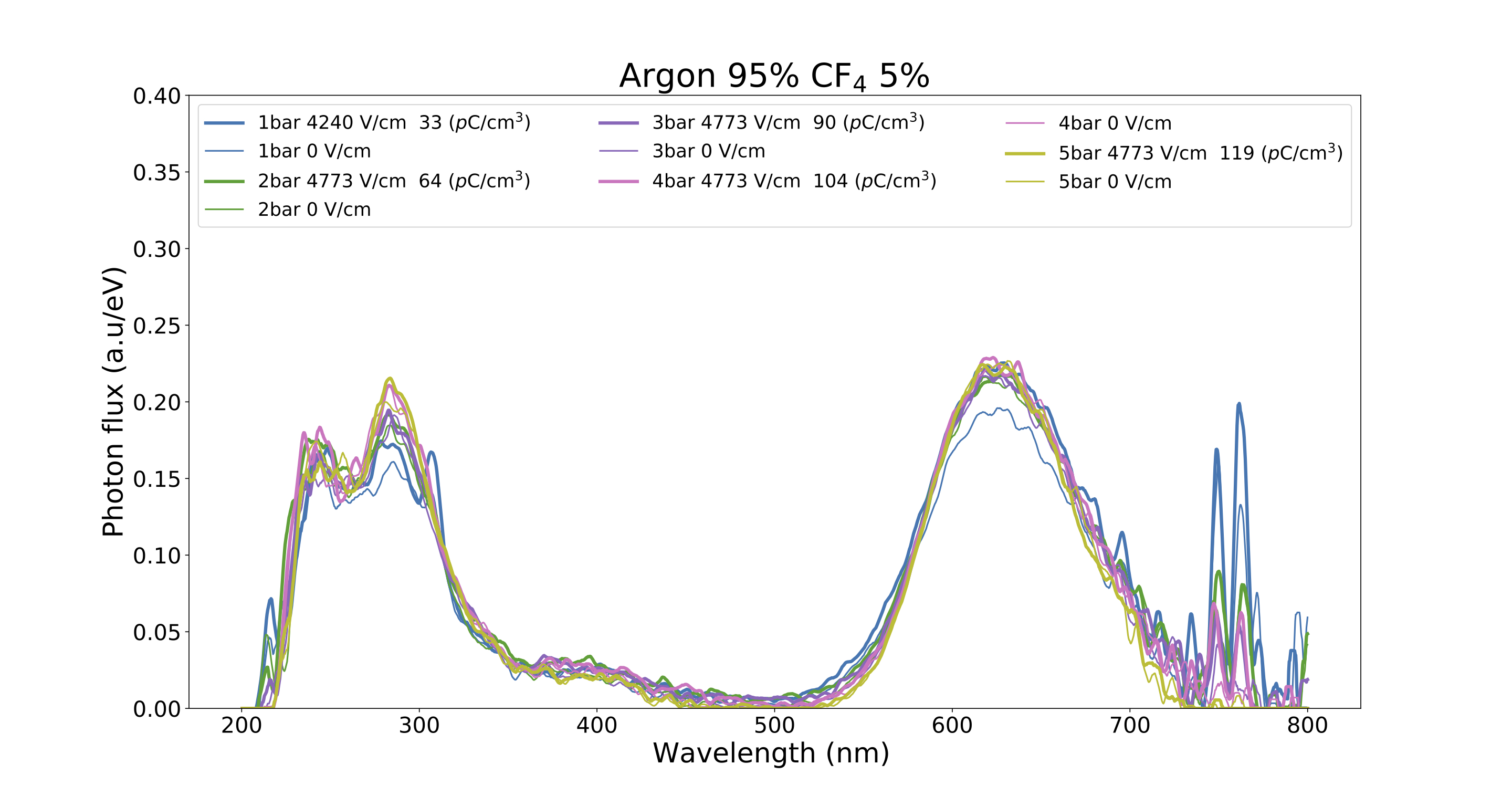}\hfill
\hspace*{-1.5cm}\includegraphics[width=1.2\textwidth]{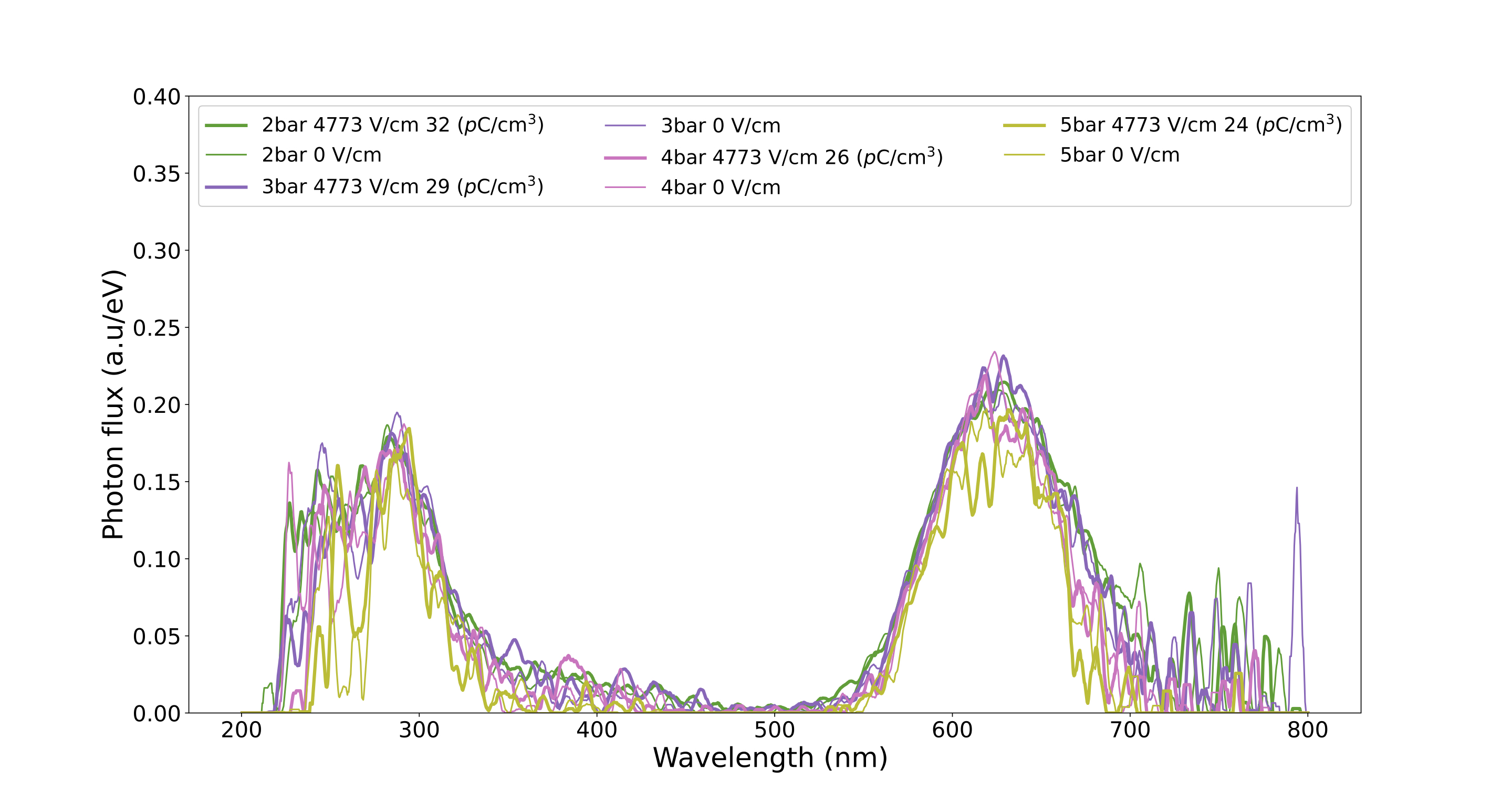}\hfill
\caption{Emission spectra per eV of released energy for Ar/CF$_4$ admixed at 95/5 and different pressures. Two different electric field conditions are considered: i) zero field and ii) a field sufficient to reach full charge-collection. Top spectra were taken with a tube current of 40~mA while bottom spectra were taken with lower tube currents of 20, 13, 10 and 8~mA, for the same irradiation time.}
\label{recombination_light}
\end{figure}

For both pure Ar and pure CF$_4$, the existence of recombination light at pressures slightly above atmospheric is well documented in case of $\alpha$-particles \cite{Ar_reco, Ar_reco2, CF4_alphas_Moro_field}: for argon, recombined electron-ion pairs landing in the first p-multiplet yield VUV photons with 100\% efficiency; for CF$_4$ the visible emission is enhanced, yet the detailed pathways are not known. In this latter case, if considering for reference the scintillation yields of pure CF$_4$ in the 500-800~nm band (796~ph/MeV for $\alpha$ particles \cite{CF4_alphas_Moro}) and its W$_I$ value of 34.3 eV/e$^-$ \cite{Reinking}, the ratio of electrons to photons is about 37. Therefore, even a 2.5\% fraction of recombined charge may double the scintillation yields if recombination light would be emitted with 100\% probability. It is thus pertinent to study the spectrum of emission for different electric fields and ionization rates. We have chosen for illustration Ar/CF$_4$ admixed at 95/5 in Fig. \ref{recombination_light}, but similar conclusions can be drawn for other admixtures. Besides the large independence observed both with electric field and ionization density, an important additional result is apparent, that has been discussed in the main text in full: wavelength-shifting properties of Ar/CF$_4$ mixtures are largely pressure-independent, in the range 1-5~bar. 

In order to better understand the absence of charge-recombination effects in our measurements it is possible to resort to previous data, e.g, considering measurements done for $\alpha$-particles in pure CF$_4$ at 1~bar \cite{CF4_alphas_Moro}, for which experimental conditions overlap. The authors reported no recombination light at any electric field in that situation. In our setup the ionization is uniform and volume recombination can be expected to be the main source of charge recombination. When comparing with the conditions in \cite{CF4_alphas_Moro}, it is thus natural to consider an uniform ionization channel too, with the average ionization and spatial extent of an $\alpha$ particle. We take for simplicity the direction along the field, although this assumption is not critical at the fields discussed (diffusion is largely isotropic). Under the prescribed-diffusion approximation \cite{JAFFE}, ionization density for $\alpha$ particles can be approximated by:

\begin{equation}
\left(\frac{dN}{dV}\right)_{\alpha} = \frac{\varepsilon}{W_I} \frac{ n^2}{(\pi R_\alpha D_T^2)}
\label{alpha_density}
\end{equation}
where $\varepsilon=5.5$~MeV is the energy of the $\alpha$ particle, $R_\alpha$ its range at $P_0=1$~bar and $D_T$ the transverse diffusion at $P_0=1$~bar for the corresponding field. The $P$-scaling factors are absorbed in the term $n=P/P_0$. Considering $R_\alpha$ = 1.6~cm \cite{Ziegler} and $D_T$= 55 \textmu m/$\sqrt{\textnormal{cm}}$ -obtained with Pyboltz for pure CF$_4$ at 1~bar- and the electric field conditions in \cite{CF4_alphas_Moro_field} (2.2~kV/cm), the average ionization density amounts to $\sim 190$~pC/cm$^3$. This is 4 times larger than in our 1~bar measurements, and about the maximum levels throughout our work (Fig. \ref{CF4_1bar_currents}-right), thus consistent with the fact that no recombination effects are seen in neither case. At 2~bar, where the average ionization density of an $\alpha$-track in \cite{CF4_alphas_Moro} would quadruple the maximum space charge densities explored here, the authors were still able to obtain a recombination-free spectrum upon application of a field of 2.2~kV/cm, comparable to our full-collection field. Conditions corresponding to $P \ge 3$-5~bar, where recombination became severe in \cite{CF4_alphas_Moro_field} (i.e., full charge-collection not being possible even upon application of fields in the few-kV/cm range), would therefore exceed the ionization densities explored here by a factor of up to 25.

\section{Gas mixing and calibration}

\begin{figure}[h!!!]
\includegraphics[width=.66\textwidth]{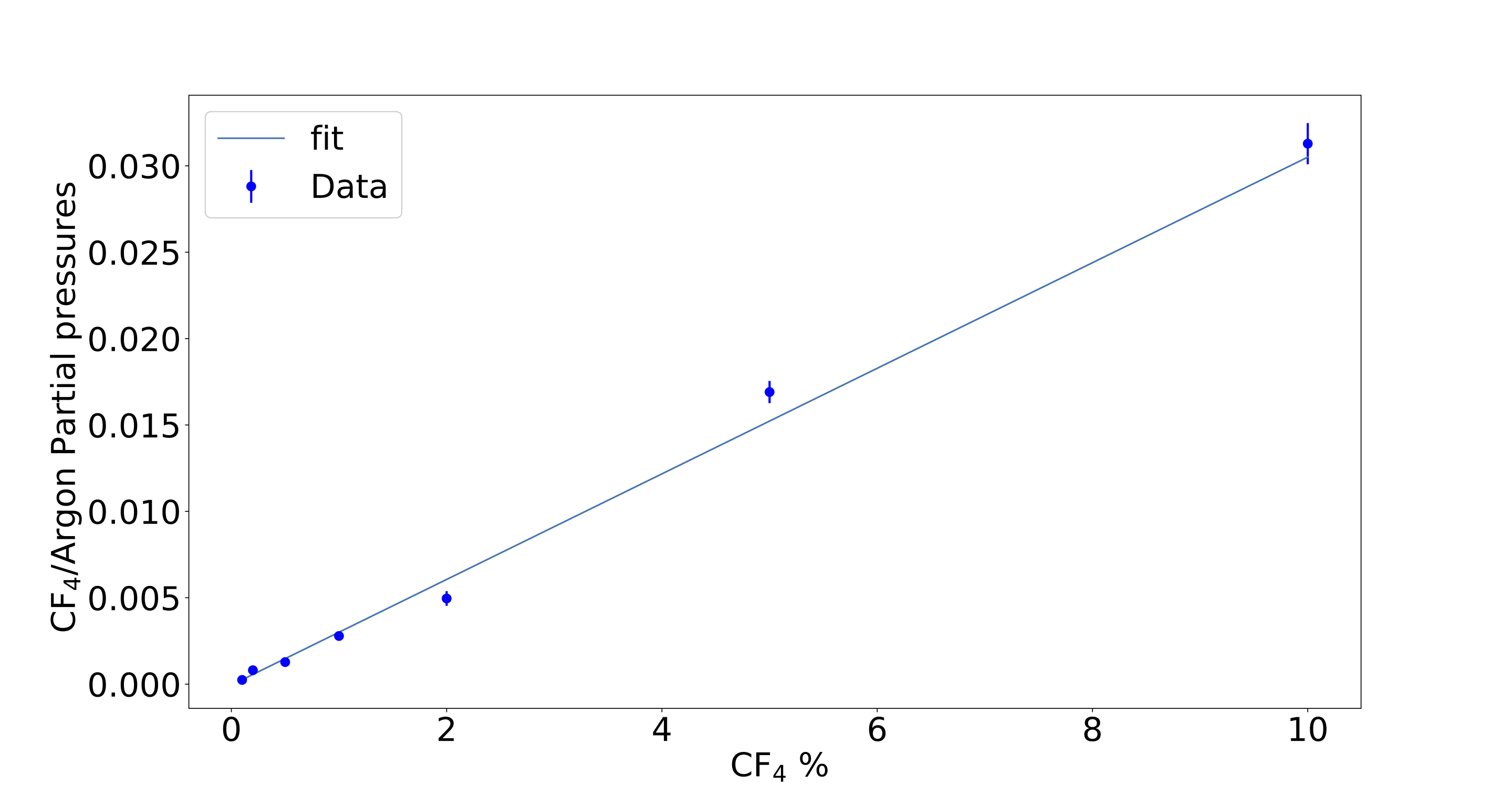}\hfill
\includegraphics[width=.33\textwidth]{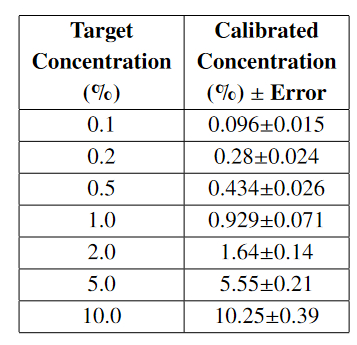}\hfill
\caption{Left: ratio of the partial pressure of CF$_4$ to argon peaks measured in the RGA during operation, as a function of the target CF$_4$ concentration. The error bar is estimated from the statistical deviation of measurements performed at different pressures and from systematic uncertainties due to the $P$-gauge resolution and dilution procedure. Right: target and calibrated concentrations, alongside their uncertainties.}
\label{CF4_RGA}
\end{figure}

Mixtures were prepared by filling at high CF$_4$ partial pressure in order to increase the precision of the pressure readings, and diluted then in argon down to low concentrations. The system was left for some time until the concentration displayed by the RGA reached a plateau, moment at which the measurements were conducted. As shown in Fig.\ref{CF4_RGA}, the pressure ratios of the main CF$_4$ and argon peaks obtained from the RGA reading show a proportional trend with the target CF$_4$ concentration estimated from the mixing procedure described above, as expected. Systematic uncertainties due to the resolution of the $P$-gauge and dilution procedure, and pure statistical uncertainties, were used to calculate the total uncertainty for each data point. A $\chi^2$-fit to a proportionality law was used on Fig.\ref{CF4_RGA} to derive a corrected value for the concentration, with its uncertainty (see adjacent table). Even if, for clarity, the target concentration has been used throughout the text, the experimentally-determined one with its uncertainties will be used in trend plots.

\end{document}